\documentclass[referee,useAMS,twocolumn,usenatbib,usegraphicx]{mn2e}

\usepackage{graphicx}
\usepackage{amssymb}
\usepackage{textcomp}
\usepackage{Times}
\usepackage{array}
\usepackage{multirow}
\usepackage{color}

\title[GRASIL-3D: Dust in Simulated Galaxies]{GRASIL-3D: an Implemention of Dust Effects in the SEDs of Simulated Galaxies}

\author[Dom\'{\i}nguez-Tenreiro et al.]{R. Dom\'{\i}nguez-Tenreiro$^{1}$\thanks{E-mail:rosa.dominguez@uam.es}, 
A. Obreja$^{1}$, G. L. Granato$^{2}$, A. Schurer$^{3}$, 
\newauthor P. Alpresa$^{1}$,  L. Silva$^{2}$,   C. B. Brook$^{1}$   and  A.  Serna$^{4}$ \\
$^{1}$Depto. de F\'{i}sica Te\'orica, Universidad Aut\'onoma de Madrid, E-28049 Cantoblanco Madrid, Spain\\ 
$^{2}$Osservatorio Astronomico di Trieste, INAF, Via Tiepolo 11, I-34131 Trieste, Italy\\
$^{3}$School of GeoSciences, University of Edinburgh, Grant Institute, The King's Buildings, West Mains Road, Edinburgh EH9 3JW, UK\\
$^{4}$Depto. de F\'{i}sica y Arquitectura de Computadores, Universidad Miguel Hern\'{a}ndez, E-03202 Elche, Spain\\
}

\begin{document}

\date{Accepted XXXX . Received XXXX; in original form XXXX}

\pagerange{\pageref{firstpage}--\pageref{lastpage}} \pubyear{2010}

\maketitle

\label{firstpage}

\begin{abstract}

We introduce  a new  model for the spectral
energy distribution of galaxies, GRASIL-3D, which includes a
careful modelling of the dust component of the interstellar
medium. GRASIL-3D is an entirely new model based on the
formalism of an existing and widely applied spectrophotometric model, GRASIL, but
specifically designed to be interfaced with galaxies with any
arbitrarily given geometry, such as galaxies calculated by
theoretical hydrodynamical galaxy formation codes. GRASIL-3D is
designed to separately treat radiative transfer in molecular
clouds and in the diffuse cirrus component. The code has a
general applicability to the outputs of simulated galaxies,
either from Lagrangian or Eulerian hydrodynamic codes. As an
application, the new model has been interfaced to the {\tt
P-DEVA} and {\tt GASOLINE} smoothed-particle hydrodynamic
codes, and has been used to calculate the spectral energy
distribution for a variety of simulated galaxies  from UV to sub-millimeter
wavelengths,
  whose comparison with observational data gives encouraging results.
In addition, GRASIL-3D  allows 2D images of such galaxies to be obtained, at several angles and in different bands.

\end{abstract}


\section{INTRODUCTION}
\label{intro}

The investigation of the process of galaxy formation and evolution by means of
hydrodynamical simulations, though not yet producing a well defined and unique
picture \citep[see for instance][]{Scannapieco:2012}, has
nevertheless already provided fundamental insights. A basic 
 limitation is that most of the processes driving the evolution of
luminous matter, which has also some back-reaction on dark matter (DM), occurs
many orders of magnitude below the resolution of any feasible cosmological
simulation, and are also poorly understood. They are implemented in the
simulation by means of approximate and uncertain prescriptions, containing
several adjustable parameters. In any case, when compared to the alternative
approach followed in the literature to understand the origin of galaxy
populations, namely the so called Semi Analytic Models (SAMs), simulations
provide, at least in principle, information on the 6D phase space as a function
of time, i.e.\ a detailed dynamical information. They also provide, for each
particle or volume element depending on the implementation, ages of the stellar
component and temperature of the gaseous one, and, when metal enrichment is
implemented, chemical composition of both.

However, to compare this rich information with the huge data
sets available nowadays,  the quite
delicate and complex step of predicting the multi-wavelength
Spectral Energy Distributions (SED) of mock objects is
required. Indeed, while simulations by themselves only trace
the evolution of mass, observations trace light. The huge
amount of multi-wavelength data collected in the last decades
have evidenced a fundamental complication and uncertainty of
this step, namely the significant reprocessing of light emitted
by primary sources (stars or active galactic nuclei) by means
of the dusty Inter Stellar Medium (ISM). Due to this effect,
which tends to be increasingly important in galaxies
characterized by higher specific star formation
rates, the predicted SED has a strong dependence on the relative
geometry of stars and dust, as well as on the optical
properties of dust grains. Both on theoretical as well as on
empirical ground the latter are expected to vary from galaxy to
galaxy, and are  difficult to predict \citep[e.g.][]{Calura:2008, Schurer:2009, Rocca:2013}.

Therefore to test the simulations against observations,
it is essential to interface their outputs with tools that can predict a
multi-wavelength SED, including a careful treatment of the
radiation transfer through the dust. A few tools 
 already exist for this  purpose,  for example:  SUNRISE
\citep{Jonsson:2004, Jonsson:2006, Jonsson:2009}; RADISHE
\citep{Chakrabarti:2008, Chakrabarti:2009}; Art2
\citep{Li:2007, Li:2008, Yajima:2012}, all using Monte Carlo
techniques to follow the radiation of photons  through  the
diffuse ISM and to calculate  a global radiation field, and
hence  the  dust re-emission. In addition  SUNRISE  
includes the treatment of star-forming  regions using the dust
and photo-ionization code MAPPINGSIII \citep{Groves:2008}.

RADISHE is a self-consistent three-dimensional code to solve
radiative transfer under the assumption of radiative
equilibrium, using a Monte Carlo code based on the
\citet{Lucy:1999} algorithm. Dust is assumed to be in thermal
equilibrium with the radiation field, and the energy it absorbs
is re-emitted as thermal emission. Therefore, the effects of
the stochastically heated small grains on the SEDs are not
considered. 

ART$^2$ also uses a Monte Carlo technique to solve the radiative transfer
in dusty media under the assumption of radiative equilibrium,
adding two modules that couple the continuum and Ly$\alpha$  line emissions, 
and take into account the effects of
ionization and dust absorption in the propagation and
scattering of photons. The Continuum module of ART$^2$ was
developed by \citet{Li:2008}, who adopted the radiative
equilibrium algorithm by \citet{Bjorkman:2001}. The new ART$^2$
version includes the treatment of Ly$\alpha$ radiative transfer
in dusty and ionized ISM as its particular improvements.

In a somewhat different context, the dust radiative-transfer
model GRASIL \citep[][ hereafter S98 and S99
respectively]{Silva:1998,Silva:1999} has been used highly
successfully for many years,  in two main ways: i), In
combination with a simple chemical evolution model, {\tt
CHE$\_$EVO}, in order to calculate SEDs,  which have then been
used to study individual  galaxies,  inferring  galaxies
properties such as star-formation histories  and dust masses
\citep[e.g.][]{Panuzzo:2007a, Calura:2008, IglesiasP:2007,
Vega:2008, Lofaro:2013}, and ii), in combination with more
sophisticated galaxy  evolution models, particularly SAMs of
galaxy formation \citep[e.g.][]{Granato:2000, Granato:2004,
Baugh:2005, Lacey:2008, Cook:2009, Fontanot:2009, Silva:2012},
in order  to predict  SEDs  which  can  be compared  to
observed ones, thereby  testing  the proposed galaxy formation
scenarios.
The GRASIL  code  has  been  used successfully with
the GALFORM model \citep{Cole:2000, Granato:2000, Baugh:2005,
Lacey:2008}, the MORGANA model \citep{Monaco:2007,
 Fontanot:2008, Fontanot:2009}; and  the  ABC model \citep{Granato:2004, Silva:2005, Lapi:2006, Cook:2009}.
It has been the first model to take into account the
age-dependent dust reprocessing of stellar populations, arising
from the fact that younger stars are associated with denser ISM
environments. GRASIL can use the outputs from SAMs, in
particular star formation and chemical enrichment histories, as
well as the limited geometric information they provide, in
order to produce self-consistent SEDs of mock galaxies
relatively quickly. Moreover, a substantially faster version of
the code was presented recently \citep{Silva:2011, Silva:2012},
exploiting artificial neural networks, which leads to improved
usability in this area. However, despite the many strengths, it
assumes equatorial and axial symmetry for the galaxy. While
this is highly suitable for SAMs, which do not calculate a more
detailed spatial distribution of stars and gas explicitly, the
use of GRASIL in coupling with galaxies produced by
hydro-simulations would imply the loss of a large quantity of
useful information.

This paper presents a new model, GRASIL-3D, based on the GRASIL
formulism, but designed to be applied to any hydrodynamic
simulation. The motivation for this work comes from adapting in
our new code the main concepts of an established code, GRASIL,
which has already proved so successful in describing galaxies
of many different types, thereby building on its strengths.

GRASIL solves the radiative-transfer equation under the
condition of thermal equilibrium for dust grains bigger than a
given size $a \ge a_{flu}$ (usually taken to be 250 \AA), with
a size-dependent temperature. However, for  smaller  grains  a
single gray body spectrum  has been found not to work
correctly. Indeed, as first  noted  by \citet{Greenberg:1968},
small grains can be stochastically heated  to temperatures much
higher than the  temperature that they would be expected to
reach if they were in temperature equilibrium. Thus a more
detailed calculation  for smaller grains  is required, as in
the \citet{Guhathakurta:1989} method incorporated in GRASIL.
This allows a proper treatment of small grains and of
polycyclic aromatic hydrocarbons (PAHs) features dominating the
mid-infrared (MIR) in some cases.

The paper is organized as follows: in $\S$\ref{CalSedSim} we
describe the GRASIL-3D model and how we interface it with the
outputs from hydro codes. Tests of the new code are presented
in $\S$\ref{TestGra3D}, and in $\S$\ref{ApplPot} some
applications to calculate the SEDs, flux density ratios, colors
and images of simulated galaxies at different stages of
evolution are discussed. In $\S$\ref{ParDiscu} the effects of
model parameter variations within their allowed ranges are
analyzed and discussed. Finally, in $\S$\ref{SummConclu} we
summarize and discuss the main results of this paper and
present its main conclusions.

\section{Calculating the SEDs of Simulated Galaxies}
\label{CalSedSim}

To develop the GRASIL-3D code, we have followed the main
characteristics and scheme of GRASIL, which we briefly recall
here. The gas is subdivided in a dense phase (fraction $f_{mc}$
of the total mass of gas) associated with young stars
(star-forming molecular clouds, MCs) and a diffuse phase ({\it
cirrus})  where more evolved ({\it free}) stars and MCs
are placed. The
young stars leave the parent clouds in the time-scale $t_0$.
The MCs are represented as spherical clouds with optical depth
$\tau \propto \delta \, m_{mc}/r_{mc}^2$ (where $\delta$ is the
dust to gas mass ratio, $m_{mc}$ is the mass of MCs, and
$r_{mc}$ is their radius), with a central source, whose
radiative transfer trough the MCs is computed. The radiative
transfer of the radiation emerging from MCs and that from free
stars is then computed through the cirrus dust (for more details
see S98 and  S99).

The aim of GRASIL-3D is to calculate multi-wavelength SEDs and
images of simulated galaxies identified in 
simulations run with either Lagrangian or
Eulerian codes. We recall that the former follow the evolution
of particles, while the latter describe the relevant dynamical
evolution through functions of position and time. In the case
of Lagrangian codes, the particles are classified into dark,
gaseous or stellar. Each galaxy-like-object produced in
the simulations is sampled by particles of these three kinds.
However, dark matter plays no role in the determination of the
galaxy SEDs.

The main quantities that determine the SED are the star
formation history, the mass of gas and the metallicity
of stars and gas. In simple models these quantities are
relatively simple functions of time, and the information on
their spatial distributions, if present, is limited to the
scale radii of stars and gas for an analytical and symmetrical
density profile. In the case of simulated galaxies from
hydro-codes, the outputs provide detailed information on the
spatial distribution of stars, gas, possibly of their
metallicity, in addition to the age distribution of stars.
All these quantities are then output at different snapshots.

\subsection{Simulation Code Outputs}

Smoothing procedures on direct outputs, or direct outputs
themselves in the case of Eulerian codes, provide, among other
functions, the following spatial mass distributions necessary
for the SEDs:

\begin{enumerate}

\item Stellar matter distribution, $\rho_{x  *}(\vec{r},
    t)$, $x$ here specifies particular properties of the
    stellar populations, for example $x$ = $y$ (young) or
    $f$ (free).

\item Gaseous matter distribution, $\rho_{x }(\vec{r},
    t)$, $x$ here specifies particular properties of the
    gas particles (e.g.\ "mc" for the dense phase, or "c"
    for the cirrus)

\item Codes where metal enrichment is implemented, provide
    the stellar or gaseous metallicity $Z_{x }(\vec{r},
    t)$,
where again $x$ specifies particular properties
       of the gaseous or stellar particles, for example,
       "cold gas" or  "young stellar population".

\end{enumerate}
In this paper we focus on how to work with the
outputs of Lagragian codes using particles,  the adaptations
needed to work with Eulerian codes being straighforward.

  To smooth out the functions above, GRASIL-3D uses a
cartesian grid whose cell size is set by the smoothing length
used in the simulation code. The smoothed functions provide
the geometry of the different simulated galaxy components, their position
dependence  $\vec{r}$ indicating that they are expected to miss
any symmetry.

Given these outputs, we need to specify in detail how to
implement in GRASIL-3D the space distributions
$\rho_{mc}(\vec{r}, t)$ (see $\S$\ref{MCDist}), and
$\rho_{c}(\vec{r}, t)$  for the diffuse ISM (see
$\S$\ref{CirrDist}), including a dust model and a scheme for
the radiative transfer within these two components (see
$\S$\ref{RadTrMC} and $\S$\ref{RadTrCirr}).

\subsection{Implementation of the Space Distribution of the MC Component}
\label{MCDist}

For the star forming molecular clouds, we follow the main
characteristics of their modeling in GRASIL, but taking
advantage of the new possibilities provided by the detailed
outputs of the simulations.  In GRASIL the mass in
the MC component in a given object at time $t_G$, $M_{mc}$, is
set by the parameter $f_{mc} = M_{mc}/M_{gas}$, where $M_{gas}$
is the model gas mass at $t_G$. This is a free global
parameter.

 Here we {\it derive} $f_{mc} $  under the assumption that
MCs are defined by a threshold, $\rho_{mc, thres}$, and by a
probability distribution function (PDF) in the gas density
pattern  the simulation returns (see below). With these
quantities, we get the total amount $M_{mc}$ of gas in the
dense phase, and correspondingly the amount and density
field of the diffuse gas. In principle, also the density field
for the dense phase of the ISM could be defined in this way, 
but here we make the assumption that all the MCs are active, i.e., they are associated
with star formation, as in GRASIL. Therefore, after defining
$M_{mc}$, we distribute it following $\rho_{y *}(\vec{r},t)$
(the density of young stars, see \ref{yslum}). This prescription is in a sense equivalent to smoothing out the subresolution MC field, 
given by the PDF, to the scale of the young stellar field, that is, to the scale of the simulation resolution, see below.
In this way, the free and young stellar field, as well as the MC field, are resolved as given by the simulation, while the diffuse gaseous field (cirrus) 
is described at subresolution scales by the PDF
\footnote{ A second possibility with only active MCs, as in GRASIL, would be to distribute the light from young stars according to the density field of MCs, 
assigning to each cell a luminous energy proportional to its MC mass content. We tested this prescription for simulated galaxy SEDs, and 
found insignificant differences with the prescriptions adopted in the text. }.

\subsubsection[]{The  probability distribution function}
\label{app1}

The PDF  is a model for the gas
    distribution at sub-resolution scales characterized by
    two parameters $\rho_0$ and $\sigma$.

Simulations at $\sim 1 $ kpc scales indicate that either
$f_{pd}(\rho) d\rho$ (i.e., the number of cells with densities in the range
$[\rho, \rho + d\rho ]$) or $f_{pd, M}(\rho) d\rho$
(i.e., the mass fraction in cells whose density is in the range
$[\rho, \rho + d\rho ]$) can be fit by log-normal functions,
characterized by a dispersion $\sigma$ (the same for both of them)
and a density parameter,
$\rho_0$ and $\rho_{0, M} = \rho_0 e^{\sigma^2}$, respectively. 
The expression for $f_{pd}(\rho) d\rho$ is:

\begin{equation}
f_{pd}(\rho) d\rho = \frac{1}{\sqrt{2\pi}\sigma} \exp [-\frac{\ln(\rho/\rho_0)^2}{2\sigma^2}] d\ln\rho
\end{equation}

The volume-averaged density $< \rho >_V$, and mass-averaged density
 $< \rho >_M$ are given, respectively, by
\citep[see, for example,][and Wada \& Norman 2007]{Elmegreen:2002}:

\begin{equation}
< \rho >_V = \rho_0 e^{\sigma^2/2}; < \rho >_M = \rho_0 e^{2 \sigma^2}
\label{rhoV}
\end{equation}

  providing a relationship between the $\rho_{0}$ and $\sigma$
parameters.

\subsubsection{Calculating $M_{mc}$}
\label{fm}

The fraction of gas in molecular clouds $f_{mc}(\vec{r},t)$ can
be calculated from a theoretical PDF, assuming that cold gas
with density $> \rho_{mc, thres}$ is in the form of MCs.
More specifically, the calculation of $f_{mc}(\vec{r},t)$  has
been made as follows:

\begin{enumerate}

\item Choose a PDF for the cold gas, $f_{pd}(ln \rho; ln
    \rho_0; \sigma)$, see Section~\ref{app1}.

\item Choose a density threshold for MC formation
    $\rho_{mc, thres}$.

\item To determine the fraction of MC mass  in the $i$-th particle at
 position $\vec{r}_i$\footnote{Note that particle positions carry a
subscript ($\vec{r}_i$), while (central points of) cell positions do not ($\vec{r}$). Otherwise,
 cells carry subscripts meaning their position in the grid ($k$-th).} 
    relative to cold gas, we calculate:

\begin{equation}
f_{mc}(\vec{r}_i; \rho_{mc, thres}, \rho_0, \sigma) = \frac{I(\rho_{mc, thres}, \rho_0, \sigma))}{I(0; \rho_0, \sigma)}
\label{fmc}
\end{equation}

where

\begin{equation}
I(\rho_{min}, \rho_0, \sigma) = \int_{\ln \rho_{min}}^{\infty} \rho f_{pd}(\ln \rho; \ln \rho_0; \sigma) d \ln \rho
\label{integral}
\end{equation}

That is \citep[][ Eq. 18 and 19]{Wada:2007}:

\begin{equation}
f_{mc}(\vec{r}_i; \rho_{mc, thres}, \rho_0, \sigma) = 0.5 (1 - Erf [z(\rho_{mc, thres}/\rho_0; \sigma )])
\end{equation}

\end{enumerate}

where $z(\rho_{mc, thres}/\rho_0; \sigma ) = (ln (\rho_{mc,
thres}/\rho_0) - \sigma^2)/ (\sigma \sqrt 2)$ and $Erf$ is the
error function. 
 The dependence on the particle position $\vec{r}_i$ is
through the parameters $\rho_0$ and $\sigma$.
To  this end, we make the identification

\begin{equation}
< \rho >_V = \rho_{gas}(\vec{r}_i)
\label{localpar}
\end{equation}

where $ < \rho >_V$ is the PDF gaseous volume-averaged density (see $\S$~\ref{app1})
and  $\rho_{gas}(\vec{r}_i)$ is the gas density of the $i$-th particle  as returned by the simulation code. 
In this way, $\rho_{0}$ can be calculated separately for each particle by combining Eqs.~\ref{rhoV} and \ref{localpar}. 
This provides a link between the two PDF (i.e., subresolution) parameters $\rho_0$ and $\sigma$  (see Eq.~\ref{rhoV}) with the (resolved) simulation output.

Then the total mass in active MCs, $M_{mc}$, can be easily
obtained through the following steps:

\begin{enumerate}

\item For each cold gas particle $i$, we calculate
    $f_{mc}(\vec{r}_i; \rho_{mc, thres}, \rho_0, \sigma) \equiv f_{mc, i}$.

\item The  mass of the $i$-th cold gas particle $m_i$ is split
    into $ m_i^{NDG} =  m_i f_{mc,i}$ for its non-diffuse
    gas content, and $ m_i^{DG} =  m_i (1 - f_{mc,i})$  for
    its diffuse gas content.

\item $M_{mc}$ is the sum of the non-diffuse gas content
of the constituent gas particles in the object.

\begin{equation}
M_{mc} = \sum_{i, object} m_i^{NDG}
\end{equation}

\end{enumerate}

\subsubsection{The MC grid-density}

In order to ascribe all the MCs to recent star formation, we have
shared out the total molecular cloud mass $M_{mc}$ in such a
way that it is proportional to the density of young stars. The
steps are the following:

\begin{enumerate}

\item Charge $\rho_{y *}(\vec{r}_i)$ to the grid to obtain
    $\rho_{y *, k}$ at  the $k$-th grid cell

\item Calculate the global MC to young star mass fraction
    in the object

\begin{equation}
\alpha = M_{mc}/M_{y*}
\end{equation}

\item The molecular cloud density at the $k$-th grid cell
    has been taken to be

\begin{equation}
 \rho_{mc, k} =   \alpha \times \rho_{y *, k}
\label{rhoMCk}
\end{equation}

As said above, the rationale behind this assignation is
that active MCs are around young stars.

\item The MC mass at the $k$-th grid cell  is:

\begin{equation}
 M_{mc, k} =   V_{k} \times \rho_{mc, k}
\end{equation}

where $V_{k}$ is the $k$-th cell volume, and must be such
that the following normalization condition holds:

\begin{equation}
\sum_{cells  k} M_{mc, k} =   M_{mc}
\label{volnorm}
\end{equation}

\end{enumerate}

\subsubsection{MCs at sub-resolution scales}

MCs sizes ($\sim 10 - 50$ pc) are smaller than the space
resolution  reached in most current hydrodynamical simulations
run in a cosmological context. Therefore, at each grid cell, a
number of MCs must be placed as follows:

\begin{equation}
N_{mc, k}  = M_{mc, k}/m_{mc}
\label{numMCcell}
\end{equation}

\subsubsection{Dust content of MCs}

Once we have the MC space distribution, we have to assign to
each MC a dust content.
We recall that a GRASIL input parameter is the dust to gass
mass ratio $\delta$, that can be set proportional to the
metallicity. We make the same assumption here. We have charged
the grid with the metallicity of the gas particles, and then,
once we know at the $k$-th cell the (cold) gas density or
mass, and the gas metallicity, we can calculate its dust
content:

\begin{equation}
\delta(Z_{k}) = \frac {Z_{gas, k}}{110 \times Z_{\odot}}
\label{dustfrac}
\end{equation}

where $Z_{gas, k}$ is the gas metallicity at the
$k$-th grid cell, calculated from the metallicities of
cold gas particles.

\subsubsection{Young Star Luminosities}
\label{yslum}

The next step is to provide the luminosity of the young stellar
populations placed inside each active molecular cloud.

As in GRASIL, we adopt the following parametrization for the
fraction $f(t)$ of the stellar populations energy radiated
inside MCs as a function of their age:

\begin{equation}
f(t) = \left\{ \begin{array}{ll}
 1         & \textrm{  $t \le t_0$} \\
 2 -t/t_0   & \textrm{ $t_0 < t \le 2 t_0$ } \\
 0         & \textrm{$t > 2 t_0$}
               \end{array} \right.
\label{fracyoung}
\end{equation}

where $t_0$ is a free parameter  setting the fraction of light
that can escape the starburst region and mimics MC destruction
by young stars ($\sim 10^6 - 10^7$ yr).\\

We calculate the SED,  $L_{\nu}^{y*}(t_i, \vec{Z}_i,
\vec{r}_i)$, for each young stellar particle ($ i =
1,...,N_{young, *}$) placed at $\vec{r}_i$, of age $t_i$,
metallicity $\vec{Z}_i$, mass $m_i$ and given IMF.
 \citet{Bruzual:2003} models were used to calculate stellar emissions.
To be consistent with hydrodynamic simulation codes (see $\S$~\ref{HydroCode}), we used  a \citet{Salpeter:1955} IMF for 
simulated galaxies identified in P-DEVA runs,  and  a \citet{Chabrier:2003} IMF for those identified in {\tt GASOLINE} runs.
Next, the luminosity at grid cell $k$ is charged
from these luminosities at particle positions.


\subsection{Space Distribution of the Cirrus Component}
\label{CirrDist}

According to the previous section, the diffuse mass content of
the $i$-th cold gas particle is given by

\begin{equation}
 m_i^{DG} =  m_i (1 - f_{mc,i})
\end{equation}

where $f_{mc,i} \equiv f_{mc}(\vec{r}_i; \rho_{mc, thres},
\rho_0, \sigma)$ is given by Eq. \ref{fmc} above, and
$\vec{r}_i$ is the $i$-th particle position.

Therefore, the  diffuse gas density  associated
to the $i$-th gas particle is:

\begin{equation}
\rho_{c}(\vec{r}_i) = (1 - f_{mc,i}) \times \rho_{gas}(\vec{r}_i)
\label{rhoDGi}
\end{equation}

This density is used to charge the grid and obtain the
diffuse gas density at the  $k$-th grid cell, $\rho_{c, k}$

To go from diffuse gas density to diffuse dust density at the
$k$-th grid cell, we use:

\begin{equation}
\rho_{dust, k}  = \delta(Z_{k}) \times \rho_{c, k}
\label{CIRRdustk}
\end{equation}

where $\delta(Z_{k})$ is as in Eq. \ref{dustfrac}.

\subsection{Dust model}   
\label{DustCompPro}

The dust  is assumed  to consist  of a mixture  of carbonaceous
and  silicate spherical grains, and PAHs. We used the optical properties of the
grains, i.e., the absorption and scattering efficiencies
$Q_{abs}$ and $Q_{sca}$ of graphite and silicate grains of
different size, computed by B.T. Draine for 81 grain sizes from
$0.001$ to $10 \mu$m in logarithmic steps $\Delta \log a =
0.05$, and made available via anonymous ftp at {\it
astro.princeton.edu}. These have been computed using Mie
theory, the Rayleigh-Jeans approximation and geometric optics as
described in \citet{Laor:1993}.\\

The dust mixture used for the diffuse ISM is that proposed by
\citet{Weingartnera:2001}. They provide a functional form for
the size distribution, and assume that when the graphite grains
are smaller than $0.01 \mu$m they take the form of PAH
molecules. The PAHs  in the diffuse cirrus consist  of a
mixture of neutral and ionized particles, the ionization
fraction depending on the gas temperature, the electron  density
and  the ultraviolet field \citep{Weingartnerb:2001}. In this
work the ionization fraction  suggested  by \citet{Li:2001} is
followed which was estimated to be an average balance for the
diffuse ISM of the Milky Way \citep[see more details in][]{Schurer_PhD:2009}.

For the size distribution of the dust grains within the dense
molecular clouds we have adopted the same composition as that
used by the GRASIL code originally described in S98. The MCs
within  the  GRASIL code have been shown to give good fits to
large star forming regions (S98), and it has been used
successfully to fit actively star-forming galaxies and
ultra-luminous infrared galaxies (ULIRGS), which are thought to
be dominated by molecular clouds (see S98 and  in particular
\citet{Vega:2005, Vega:2008, Lofaro:2013}). It is also
important to note that the abundance of PAHs in molecular
clouds have been specifically tuned  by \citet{Vega:2005}, hereafter V05, to
agree with the MIR properties of a sample  of local actively
star-forming galaxies. Due to the  extensive testing  that the
molecular clouds within GRASIL have received, the adoption of
the same dust composition should therefore  represent an
excellent choice for inclusion in this work.

We also recall that both dust distributions are calibrated on
the same observables, i.e. the average extinction curve and
cirrus emission in the Milky Way.

Once the size distribution and the optical  properties for the
dust mixture have been set, it is then  possible to calculate
its absorption, scattering and extinction optical depth (see
S99 for a summary of the definitions).


\subsection{SED Determination with GRASIL-3D}

The aim of the model is to calculate the radiant flux (luminosity)  from a given object measured by an
external observer in a given direction  $(\theta,\phi)$ and at wavelength
$\lambda$, using the expression (see S98 for more details):

\begin{equation}
F_{\lambda}(\theta,\phi) = 4 \pi \sum_k V_k j_{\lambda, k} \exp[-\tau_{eff, \lambda}(k;\theta,\phi)]
\label{obsflux}
\end{equation}

(with units erg s$^{-1}$ \AA$^{-1}$), and where the sum is over the different small volumes $V_k$ (the
grid) over which the object has been divided, and

\begin{equation}
j_{\lambda, k} = j_{\lambda, k}^{mc} + j_{\lambda, k}^{*} + j_{\lambda, k}^{c}
\label{obsflux2}
\end{equation}

is the volume emissivity (erg cm$^{-3} $s$^{-1}$ \AA$^{-1}$ sr$^{-1}$) of the k-th
volume element at wavelength $\lambda$, with $mc, * $ and $c$
corresponding to molecular cloud, free stellar and cirrus
components, and $\tau_{eff, \lambda}(k;\theta,\phi)$ is the
effective  optical thickness for cirrus absorption from the
$k$-th volume element to the outskirts of the galaxy along the
$(\theta,\phi)$ direction. We describe these terms in the following sections,
separately for MCs and cirrus, for which we solve the radiative transfer with
different methods. Details on the computation of the dust
emissivity, both for grains in thermal equilibrium and
fluctuating ones, can be found in S99.

\subsubsection{Radiation transfer in MCs}
\label{RadTrMC}

The radiation transfer through the molecular clouds within
GRASIL-3D is calculated using the same technique as in GRASIL,
since MCs have the same characteristics as those considered
by S98, where the starlight emitted from within a MC is
approximated as a central source and MCs are spherical. We
summarize here only the main features which are most important to the
implementation in the new code.

The  radiative transfer is solved using the
\citet{Granato:1994} code, with the $\lambda$-iteration method,
i.e. at each successive iteration the local temperature of the
dust grains is calculated from the radiation field of the
previous iteration. In such a way the code converges to a value
for the radiation field at all radii of the molecular cloud
which will give the correct dust temperature.

This simplified geometry results in a considerable decrease in
the computational time. However it is insufficient to match the
complex system  of randomly distributed hot spots and cooler
regions  observed  in real star forming molecular clouds, and
could lead to unrealistically hot dust spots in their center. A
maximum  inner edge temperature was introduced in S98 to
compensate for this potentially too hot temperature. This was
shown to be sufficient for the modelling of molecular clouds,
giving good fits to observed data.

Given the central stellar source, the SED emerging from MCs
depends only on one parameter, their optical depth:

\begin{equation}
\tau_{mc} \propto \delta \frac{m_{mc}}{r_{mc}^2}
\label{taumc}
\end{equation}

Since we set $\delta \propto Z$ (Eq. \ref{dustfrac}), and $Z$
now is a local quantity, the value of $\tau_{mc}$ depends on 
the cell position. 
This means that we have to calculate the radiation transfer  separately for the MCs in each grid cell, due to the different values of $\tau$ and the central stellar
source (see \ref{yslum}), even if we set the same $t_0$ for
all MCs.

\subsubsection{Radiation transfer through the diffuse cirrus}
\label{RadTrCirr}

The same two assumptions as in GRASIL have been introduced to simplify the 
radiative transfer through the diffuse dust, namely:

\begin{enumerate}

\item The effect of dust  self-absorption  is ignored.

\item The  effect of UV-optical  scattering is approximated by means  of an effective optical depth,  given by the  geometrical  mean  of the  absorption and  
scattering efficiencies \citep{Rybicki:1979}:

\begin{equation}
\tau_{eff, \lambda}^{2} = \tau_{abs, \lambda} (\tau_{abs, \lambda} + \tau_{sca, \lambda}).
\label{difopdep}
\end{equation}

\end{enumerate}

These approximations have been shown to give similar results
when compared  to the  more rigorous Monte Carlo techniques of
\citet{Witt:1992} and \citet{Ferrara:1999} in the majority of
cases tested (see S99).
Using the  two assumptions stated
above, the local (angle averaged) radiation field in the $i$-th
grid cell $J_{\lambda, i}$ (units erg s$^{-1}$ \AA$^{-1}$ sr$^{-1}$ cm$^{-2}$) due to the extinguished emissions of
the free stars and molecular clouds from all the other  cells
can be calculated using the same equations as in the original
GRASIL code (see S98 and S99), namely:

\begin{equation}
J_{\lambda, i} = \sum_k V_k (j_{\lambda, k}^{mc} + j_{\lambda, k}^{*}) \times \exp[-\tau_{eff, \lambda}(i,k)] \times r(i,k)^{-2}
\label{RadField}
\end{equation}

where the sum is over the different cells of volume $V_k$ over
which the simulated galaxy has been divided.
$j_{\lambda,k}^{mc}$ and $ j_{\lambda, k}^{*}$ are the volume
emissivity of the $k$-th volume element at wavelength
$\lambda$, with $mc$ and $*$ corresponding to molecular cloud
and free stellar components, respectively, and $\tau_{eff,
\lambda}(i,k)$ and $r(i,k)$ are the effective optical thickness
and distance from the $i$-th to the $k$-th grid cells.
Once the radiation field $J_{\lambda, i}$ has been calculated
at any cell $i$, the cirrus emissivity $j_{\lambda, k}^{c}$ is
calculated following S98, section 2.4.

Together with the MC and free stars emissivity, the cirrus
emissivity is used to calculate the emerging SED,
$F_{\lambda}(\theta,\phi)$ (Eqs. \ref{obsflux} and
\ref{obsflux2}).\\

A concern is in order when calculating the radiation field
within the $i$-th cell caused by the stellar and MCs emissions
within the same cell. In this case a (apparent) singularity
appears. To overcome this problem, the cell is split
into $N_p$ random points, $P_n$, each representing a
small volume $V(P_n) = V_i/N_p$, instead of being represented
by its center, and the radiation field in the $i$-th cell
caused by its own emission, $J_{\lambda, i; i}$, is calculated
as:
\begin{equation}
J_{\lambda, i; i} = \sum_{n=1}^{N_p} \frac{J_{\lambda, i} (P_n)}{N_p}
\end{equation}

where $J_{\lambda, i} (P_n)$ is the radiation field at $P_n$ caused by
 the emission from the small volumes $V(P_m)$ represented by the remaining
$N_p -1$ random points:

\begin{equation}
J_{\lambda, i} (P_n) =  (\frac{V_i}{N_p}) (j_{\lambda, i}^{mc} + j_{\lambda, i}^{*}) \sum_{m=1}^{N_p -1} \frac{ \exp[-\tau_{eff, \lambda}(m,n)]}{ d(m,n)^{2}}
\label{ParRF}
\end{equation}

where again $j_{\lambda, i}^{mc}$ and  $j_{\lambda, i}^{*}$ are
the volume emissivity at the $i$-th cell coming from molecular
clouds and stars, respectively; $d(m,n)$ is the distance
between the $m$-th and the $n$-th random points at cell $i$;
and

\begin{equation}
\tau_{eff, \lambda}(m,n) =  \sigma_{\lambda} \times n_{H,i} \times d(m,n)
\label{taumn}
\end{equation}

is the effective optical
thickness between these $m$-th and the $n$-th random points.

It is worth noting than when max$_{ith cell}$ $\tau_{eff, \lambda}(m,n)$ is
$<< 1$, then the following approximation holds:

\begin{equation}
J_{\lambda, i; i} = V_i \times (j_{\lambda, i}^{mc} + j_{\lambda, i}^{*}) \sum_{n=1}^{N_p} \sum_{m=1}^{N_p -1} d(m,n)^{-2}/N_p^2
\end{equation}

where the average of the inverse squared distances $d(m,n)$ is 0.42 when $P_n$ and $P_m$ points
cover the unit cube.
When taking this average, a volume-like factor going as  $d^2$ appears such that
no singularity is present when $d \rightarrow 0$.

However, when the effective optical thickness within the $i$-th
cell is not $<< 1$, then we have a double summation involving
$\tau_{eff, \lambda}$ to be calculated through the Monte Carlo
cell splitting. 
Furthermore,  in these situations,
determining the contribution of the 26 neighboring cells to
$J_{\lambda, i}$ using just the central points of the 27 cells
could lead to inaccuracies in the results too. Therefore, a similar
treatment was also applied to these 26 cells by splitting them
into random sub-volumes.

For the practical implementation, the calculations
have been made in the unit cube (i.e., cell side $L$=1). 
As  $n_{H,i}$  
in the expression giving $\tau_{eff, \lambda}(m,n)$ in Eq.~\ref{taumn} above does not change in the cube, 
the results can be rescaled to the actual grid
cell side $L \neq 1$. 
To this end, the $L$=1 results have been tabulated for different values of $\tau_{eff, \lambda}(m,n)$. 
This allows calculation of  the splitting averages (see
for example Eq.~\ref{ParRF}) 
to be performed only once, with the tabulated results then applied to all of the different cells.
Therefore, cell splitting does not add
CPU time to the calculations.

\subsubsection{Radiant Flux Calculation}
\label{SpeFluxCal}

Also the practical computation of the radiant flux in Eq.
\ref{obsflux} requires a correct implementation. At high
optical thicknesses, when calculating the absorption of the
radiation emitted at the $k$-th cell along a given direction
within this same cell, representing the cell by its central
point gives a poor representation of the extinction. A better
approximation is  again obtained by splitting the cell into $N_p$
random points, $P_n$, and then the  extinction is calculated
along rays emerging from $P_n$, along directions ($\theta,
\phi$), and finally calculating the averages at fixed directions
as $P_n$ covers the cell. Specifically, the contribution of the
$k$-th cell to the radiant flux can be written as (i.e., from
emissions within the $k$-th  cell):

\begin{equation}
F_{\lambda, k}(\theta,\phi) = \frac{1}{N_p} \sum_{n=1}^{N_p} F_{\lambda, k} (P_n;\theta,\phi)
\label{ParFlux}
\end{equation}

where $F_{\lambda, k} (P_n;\theta,\phi)$ is the contribution to
the emission $F_{\lambda, k}(\theta,\phi)$ of the small
volume $V(P_n) = V_k/N_p$ around point $P_n$ resulting from
cell $k$ splitting. To calculate the extinction at cell  $k$ of
rays emerging at point $P_n$, one has to calculate the
distances from $P_n$ to the cell borders along the ($\theta,
\phi$) directions, $d(n;\theta, \phi)$. Indeed, as the
emissivity is the same at any $V(P_n)$, the averages in Eq.
\ref{ParFlux} just involve the partial optical depths:

\begin{equation}
\tau_{\lambda, k}(n;\theta, \phi) = \sigma_{\lambda} \times n_{H,k} \times d(n;\theta, \phi),
\label{ParTau}
\end{equation}

namely:

\begin{equation}
\frac{1}{N_p} \sum_{n=1}^{N_p} \exp[- \tau_{\lambda, k}(n;\theta, \phi)].
\label{AvExt}
\end{equation}

In the practical implementation, as explained in $\S$\ref{RadTrCirr},
the calculations have 
been made just once and for all in the unit cell,
and the resulting  values normalized to the actual cell sizes.
Therefore, here again cell splitting does not add CPU time to
the calculations.

The emission $F_{\lambda}(\theta,\phi)$ is the key tool to
calculate simulated galaxy SEDs and their derivatives, such as
luminosities, colors and images in different bands from the UV
to the sub-mm.

\subsubsection{Calculating Images with GRASIL-3D}

GRASIL-3D allows us to calculate images of simulated objects in different filters
and as viewed from different directions ($\theta, \phi$).
To this end, the rectangular grid is oriented with the $z$-axis in the chosen direction, and then,
to calculate the radiant flux at point $(x,y)$ (i.e., in the plane normal
to the line-of-sight),  we sum:

\begin{equation}
F_{\lambda, x,y} = \sum_k  F_{\lambda,k}(\theta, \phi)
\end{equation}

where $F_{\lambda,k}(\theta, \phi)$ is the extinguished emission from the
$k$-th cell, and the sum goes over all the cells in the line-of-sight.
In $\S$\ref{ImagesResults} some examples of images of simulated objects
calculated with GRASIL-3D are shown.

\subsection{Summary: GRASIL-3D Parameters}
\label{Parameters}

The  GRASIL-3D  code  contains  several  free parameters
for transposing  particle positions from the
simulation outputs onto the  grid. These are new relative to the
GRASIL code. The  meaning  of these  parameters will be
summarized, and  the  possible range of values each can take
will be discussed in turn.

\begin{enumerate}

\item \underline{The grid size:}   A rectangular grid has
    been used.
Its size must be such that it does
    not spoil the galaxy space resolution returned by the
    simulations. Therefore, the softening parameter of the
    simulations $\epsilon$ sets the grid size.

\item \underline{Threshold density for molecular clouds:}
 Gas with densities above $\rho_{mc, thres}$ are set to
    form molecular clouds.
    A threshold value is commonly used by numerical
    simulations of molecular clouds.
    This value is backed up by a large number of
    observations of molecular clouds both from our own
    galaxy and nearby galaxies. Values for this parameter
    used  in  simulations have a range of values depending on the authors. 
    For example,  
    $\rho_{mc, thres}$ = 100 H nuclei cm$^{-3} \approx $3.3  $ M_{\odot}$  pc$^{-3}$ in \citet{Tasker:2009}
    and   $= 35$ H nuclei cm$^{-3} \approx 1 M_{\odot} 
     pc^{-3}$ in \citet{BallesterosP:1999} and
    \citet{BallesterosP_Scalo:1999}. An adequate range  for
    this parameter is therefore  taken to be $\rho_{mc,
    thres}$ = 10 - 100 H nuclei cm$^{-3}$.

\item \underline{Parameters for the log-normal PDF:} Two
    parameters $\rho_0$ and $\sigma$ govern the log-normal
    PDF function, see $\S$~\ref{app1}. They are linked to  the
    density  of gas by Eqs. \ref{localpar} and \ref{rhoV}, 
    so it is convenient to fix one of the parameters
    and use the  equations to calculate the  other.
    We fix $\sigma$, treating it as a free
    parameter, and compute $\rho_0$ from the gas density. 
    Values for $\sigma$ have been calculated to
    range from 2.36 to 3.012 in \citet{Wada:2007},
 while \citet{Tasker:2009} give $\sigma$ = 2.0.

Between them, $\rho_{mc, thres}$ and $\sigma$,
control  the calculation of  the cirrus field and the proportion of
the  total  gas in the  galaxy  in the  form of molecular
gas, $f_{mc}$. A useful check for the  choice of these parameters can
be made by comparing  the final calculated average  value
for the galaxy with observations.  For example
\citet{Obreschkow:2009} give the ratio  as a function  of
the galaxy morphological  type and gas mass (Figures 4 \& 5
in that paper), see also \citet{Leroy:2008}. 
More recently \citet{Saintonge:2011, Saintonge:2012} 
conducted COLD GASS, a legacy survey for molecular gas in
nearby, massive ($M_{star} \ge 10^{10} $M$_{\odot}$)
galaxies. Data on the molecular gas content in normal star
forming galaxies at $z \sim 1 - 3$  are being gathered and
analyzed by the PHIBSS team \citep{Tacconi:2013}, a considerable
improvement compared to our current understanding.

\end{enumerate}

Otherwise, the treatment of the  dust  properties and the MC
model closely follow the GRASIL code. We summarize the
corresponding parameters as well as their range of values.

\begin{enumerate}

\item \underline{Escape time-scale from MCs, $t_0$:} This
    parameter represents  the  time taken  for stars to
  escape the molecular clouds where they were born. $t_0$
  is likely to be of the order of the lifetime of the most
  massive stars, with masses $\sim$ 100 $M_{\odot}$ to 10
  $M_{\odot}$ and corresponding lifetimes of 3 to 100 Myrs.
  In practice the  timescale  is likely to  vary with the
  density of the  surrounding ISM. For low-density
  environments like spiral  galaxies,  with  low star
  formation rates, the lifetime is likely to be nearer the
  lower end of the  range,  corresponding  to that of the
  most massive supernovae. On the other hand,
  $t_0$ is likely to be much closer  to the longest value  for a 
  high-density environment like the star-forming central 
  regions of starburst galaxies.

The escape timescale was found to be a very important
parameter of the GRASIL model. Typical values were found by
comparison  to local observations in S98, yielding $\sim$
2.5 to 8 Myrs for normal spiral galaxies, and between 18
and 50 Myrs for strong starburst galaxies.

\item \underline{Optical depth of MCs:} As shown in Eq.
    \ref{taumc}, the optical depth of MC depends on a
    combination of $r_{mc}$ and $m_{mc}$.
Observational estimates from our Galaxy suggest typical
values $m_{mc} \sim 10^5$ to $ 10^6 M_{\odot}$ and $r_{mc}
\sim $  10 to 50 pc. In order to match observed SEDs of
local galaxies S98 set  $m_{mc} \sim 10^6 M_{\odot}$  and
derived values of $r_{mc}$ between 10.6 pc and 17 pc.

\end{enumerate}

The parameters needed to characterize the grain size
distribution have been fixed, and, therefore are not considered
as free in this work (see $\S$\ref{DustCompPro}). Otherwise, it
is worthwhile to recall that the simulations provide and fix the
geometry of each galaxy component, as well as its SFR, metal
enrichment and gas fraction histories in such a way that no further parameters
are needed to describe them.

\subsection{Numerical Performance of the GRASIL-3D Code}
\label{NumPer}

GRASIL-3D is a Fortran paralell (MPI) code. It has been run with up to 1024 CPUs at Red Espa\~nola de Supercomputaci\'on (RES), with good scaling properties. 
The optimal number of CPUs depends on various factors, of which the number of cells containing baryonic particles is the most important. 
This number can be expressed as a covering factor multiplied by the total number of cells, $L_{\rm box}^3$. While $L_{\rm box}$ is given as input according to the spatial resolution, 
the covering factor depends on the particulars of each hydrodynamical simulation.
Typical values for the runs involving HD-5103B galaxy-like object presented in this work, with 60$^3$ grid cells and a covering factor close to 1, are 7500 sec with 32 CPUs. 

GRASIL-3D can be considered a kind of {\it software telescope}, that is, a software device performing similar tasks
as a telecope does. Indeed, the device can be used in two main modes,
either to obtain images of simulated galaxies, or to obtain their SEDs and associated photometric
properties: fluxes in different bands and colors. 
  
The memory requirements depend on the using mode.
Using high resolution SSP spectra (8500 wavelengths between 91 angstrom and
10 $\mu$m), the code needs 25, 100  and 300 Mb of RAM/CPU for meshes of
60$^3$, 100$^3$,  150$^3$  cells, respectively, in the SED-only mode. In the
current imaging mode, GRASIL-3D produces 3 images corresponding to 3
lines of sights perpendicular to each other. For images,  no high resolution SSP spectra
are generally needed. Therefore, using low resolution SSP spectra (for example, 200 wavelengths between 91 angstrom 
and 10 $\mu$m), the memory requirement would be 36  and 95
Mb of RAM/CPU for meshes of 60$^3$ and 110$^3$, respectively, with these requirements scaling linearly
with the number of wavelenghts in the SSPs.


\section{Testing GRASIL-3D}
\label{TestGra3D}

\subsection{Energy Balance}
\label{EnBal}

The total energy absorbed by the MC or cirrus dust component
has to be equal to the total energy they separately emit.
Therefore, energy balance in either component is a necessary
condition to be met by GRASIL-3D. As mentioned, the radiative
transfer in MCs is as in the GRASIL model. To check the energy
balance in the diffuse ISM, we calculate its heating by stars
and MC emissions, and compare it with the total (integrated
over directions and wavelengths) diffuse ISM emission.

The total energy absorbed by the cirrus component can be
written as:

\begin{equation}
E_{abs} = \sum_k \int d\lambda  \int d\Omega E_{abs,\lambda}(k;\theta,\phi)
\end{equation}

where we have summed over the grid cells  $V_k$, and integrated
over wavelength and direction the components:

\begin{equation}
E_{abs, \lambda}(k;\theta,\phi) = V_k j_{\lambda,k}(1-\exp[-\tau_{eff, \lambda}(k;\theta,\phi)])
\end{equation}

i.e., the absorption of energy emitted at cell $k$ at
wavelength $\lambda$ along rays traveling in the
$(\theta,\phi)$ direction to the outskirts of the galaxy, where
$j_{\lambda,k}$ is the volume emissivity of the $k$-th cell
(i.e., from stellar and MC emissions), and $\tau_{eff,
\lambda}(k;\theta,\phi)$ is the effective optical depth (Eq.
\ref{difopdep}) from the (central point of the) $k$-th cell to
the outskirts of the galaxy along the $(\theta,\phi)$
direction.

Within this scheme, calculating the  energy absorbed at cell
$k$ from its own emission is not straightforward. It can be
written:

\begin{eqnarray}
\lefteqn{E_{abs}(k,k) = V_k \int d\lambda j_{\lambda,k} {} }\nonumber       \\
 & & {}  \times \int d\Omega (1-\exp[-\tau_{eff, \lambda}(k,k;\theta,\phi)]){}
\label{Ekk}
\end{eqnarray}

and involves $d(k,k;\theta, \phi)$, that is, the distance
traveled by a ray from the central point of the cell to its
border along the $(\theta, \phi)$ direction. We note that when
max$_{ith cell}$ $\tau_{eff, \lambda}(m,n)$ is $<< 1$, then

\begin{equation}
1-\exp[-\tau_{eff, \lambda}(k,k;\theta,\phi)] \simeq \sigma_{\lambda} n_{H,k}  d(k,k;\theta, \phi)
\end{equation}

and therefore the integral over directions is the average
distance from the central point of a cell to its borders, along
random directions.

When this condition is not satisfied,  representing the cell by
its central point gives poor results. 
A better approximation is
obtained by splitting the cell in $N_p$ random small volumes
represented by points $P_n$, and then taking averages as in the
previous section. 
By doing so, the energy absorbed at cell $k$
from its own emission (Eq.~\ref{Ekk}) can be written as:

\begin{eqnarray}
\lefteqn{E_{abs}(k,k) = \frac{V_k}{N_p} \int d\lambda j_{\lambda,k} {} }\nonumber       \\
& & {} \times \int d\Omega \sum_{n=1}^{N_p} (1-\exp[-\tau_{eff, \lambda}(n;\theta,\phi)]){}
\end{eqnarray}

involving the partial optical depths defined in
Eq.~\ref{ParTau}. The angle average gives, in the limit of low
optical thickness, the average distance from the central point
of a cell to its borders along random directions, as expected.
However, when this is not the case, cell splitting to calculate
the heating by its own emissions leads to important
differences. The practical implementation consists in
calculating the splitting in the unit cell, and then rescale
to the actual cell size.

The calculation of this heating allows us to test the energy
balance within the cirrus. In  Tables \ref{MassLumDisk} and \ref{PropTab5103} we give
some results  for both  the MC and the cirrus components, as well
as for the overall bolometric luminosity. We see that the
energy conservation is very good.

\begin{figure}
\centering
\includegraphics[width=.35\textwidth, angle=270]{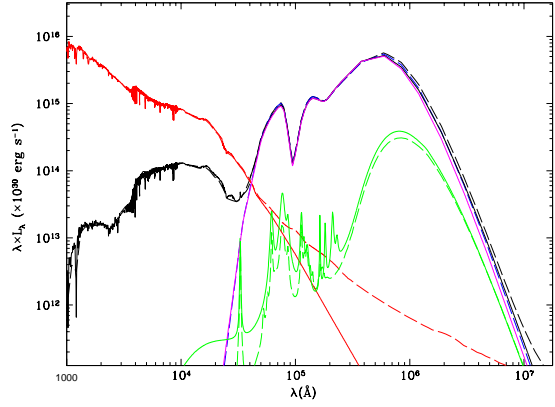}
\caption{The SEDs of two models corresponding to ARP220 with a
geometry of a King sphere. Dashed lines are for GRASIL, while
continuous lines are for GRASIL-3D  operating on a
Monte Carlo realization of the same model, see text. The black
lines are the final SEDs when dust effects are taken into
account, the dashed blue and continuos magenta lines correspond to
extinguished MCs emission (for GRASIL and GRASIL-3D, respectively), green to cirrus emission, and red to stellar
emission when the dust effect is not taken into account.}
\label{SedArp}
\end{figure}

\subsection{GRASIL-3D versus GRASIL Results}

As GRASIL-3D is based on the formulism of GRASIL 
(with slightly different dust model implementations), comparing
the SEDs they produce when applied to the same galaxies is a
necessary check for GRASIL-3D.

To this end, we have produced analytical galaxy models to be
compared with GRASIL results. They are Monte Carlo realizations
with $N_{part}$ particles of King spheres (see S98) and Monte
Carlo realizations of SFR histories. More specifically,
following S98 we have built a Monte Carlo model for ARP220 with
a core radius $r_c = 0.5$ kpc, an age of 13 Gyr, where the SFRH
$\phi(t)$ is constant in the interval of the age of the
Universe 1 Gyr $< t_u < $ 12.95 Gyr, involving a stellar mass
of $2.3 \times 10^{11} M_{\odot}$. Later on, for 12.95 Gyr
$< t_u < $ 13 Gyr it undergoes an exponential burst with
e-folding time $t_e = 0.05$ Gyr involving a gas mass of 2.5
$\times 10^{10} M_{\odot}$. The gas mass fraction at $t_u =$ 13
Gyr is $f_{gas}=$ 0.139, the fraction of gas in molecular
clouds is assumed constant at $f_{mc}$=0.5 and the parameter
regulating the escape of young stars from MC is $t_0 =$50 Myr,
with a solar metallicity and molecular cloud radii of
$r_{mc}=10.6$ pc.

In Figure~\ref{SedArp} we show the SEDs for this model,
calculated with both GRASIL-3D and GRASIL. Taking into account
the differences of the dust model implementation, that mainly
affect the cirrus and   in particular their  PAH emission (see discussion
in  V05), the agreement can be considered very
satisfactory. 
Note that the  difference in the intrinsic stellar emission appears because the reported GRASIL computation (dashed red line) 
includes the emission from dust in the stellar envelopes directly into the stellar population models, following \citet{Bressan:1998} and \citet{Bressan:2002}, 
while that of  GRASIL-3D (full red line) does not.

\section{Applications and Potentialities}
\label{ApplPot}

GRASIL-3D has been interfaced with a variety of galaxies
identified in cosmological hydrodynamic simulations. Here we
show results of galaxies run with two different codes: P-DEVA
and {\tt GASOLINE}. 
We note that we are not able to make accurate statistical analyses yet, as we do not have a statistical number of high resolution galaxy simulations (in fact, no group currently has such a set). Indeed, all we  currently can achieve is to show 
that by interfacing GRASIL-3D with simulations (whose validity has been otherwise proved, see below), we obtain results consistent with observational data.

\subsection{Hydrodynamic Codes}
\label{HydroCode}

\subsubsection{P-DEVA}

P-DEVA is an entropy-conserving AP$^3$M-SPH code, an OpenMP
parallel version of the DEVA code \citep{Serna:2003}, which
includes the chemical feedback and cooling methods described in
\citet{Martinez:2008}. The primary concern when developing this
code was that conservation laws (e.g. momentum, energy, angular
momentum and entropy) hold accurately \citep[see][for
details]{Serna:2003}. The star formation recipe implemented in
the DEVA code follows a Kennincutt~\textendash~Schmidt-like law
with a given density threshold, $\rho_*$, and star formation
efficiency $c_{*}$. In line with \citet{Agertz:2011},
inefficient SF parameters are implemented, which implicitly
account for the regulation of star formation by feedback energy
processes by mimicking their effects, which are assumed to work
on sub-grid scales.

The chemical evolution implementation \citep{Martinez:2008}
accounts for the full dependence of the radiative cooling on
the detailed composition of the gas, through a fast algorithm
based on a metallicity parameter, $\zeta(T)$, which takes into
account the weight of the different elements on the total
cooling function. The code also tracks the full dependence of
metal production on the detailed chemical composition of
stellar particles \citep{Talbot:1973}, through a Q$_{ij}$
formalism implementation of the stellar yields, for the first
time in a SPH code. A probabilistic approach for the delayed
gas restitution from stars reduces the statistical noise and
allows for a detailed study of the inner chemical structure of
objects at an affordable computational cost. Moreover, the
metals are diffused in such a way as to mimic the turbulent
mixing in the interstellar medium, and the radiative cooling
depends on the detailed gas particles metal composition.

Hydrodynamic evolution leads to gas particles being  either
cold or hot, as gas temperature distribution in simulated
galaxies has two conspicuous  maxima. Stars mostly form from the
cold component. As we have said above, the star formation
history is directly provided by the simulation. Stellar
particles are considered as SSPs, with given age, metallicity
and IMF $\Phi(M)$, specifically a Salpeter
IMF \citep{Salpeter:1955}, with a mass range of
[M$_l$,M$_u$]=[0.1,100]M$_\odot$.
According to the chemical evolution scenario
implemented in P-DEVA, stellar particles can be transformed
into gaseous ones, according to a probabilistic rule.

\begin{table*}
\begin{minipage}{6.5in}
\renewcommand{\thefootnote}{\thempfootnote}
\centering \caption{Data on the $z=0$ sample of simulated disk-like galaxies}
\begin{tabular}{l ccccccccc}
\hline
Name & $m_{\rm bar}$\footnote{Baryonic particle mass. For {\tt GASOLINE} runs, its initial value is given } & $M_{\rm star}$ &  $M_{\rm gas}$ & SFR\footnote{Integrated over the past 100 Myrs} & $<Z_{\rm star}>$\footnote{Average stellar metallicity (mass fraction)} &  $<Z_{\rm gas}>$\footnote{Average gas metallicity (mass fraction)}  & r$_e$\footnote{Bulge scale length in the $r$ band. A value 0 means pure exponential disk} & r$_s$\footnote{Disk scale length in the $r$ band} & B/D\footnote{Bulge to disk luminosity ratio in the $r$ band} \\
     & ($10^{5}$ M$_{\odot}$) & ($10^{10}$ M$_{\odot}$) & ($10^{10}$ M$_{\odot}$) &  (M$_{\odot}$/yr) &   (10$^{-2}$) & (10$^{-2}$) &  (kpc) & (kpc) &   \\
\hline
\hline
g1536\_L$^{*}$  &  1.9  &   2.32 &  1.97 &  1.809 &   1.17  &    1.25  &       1.38 &  3.86 &  0.35    \\
g21647   &  0.25  &   2.32 &  1.37 &  3.895 &   1.48  &    1.94  &       0.00  &  1.52 &  0.00    \\
g7124    &  2.00  &   0.60 &  1.10 &  0.314 &   0.46  &    0.86  &       0.00  &  2.94 &  0.00    \\
LD-5003A &  3.82 &   1.66 &  0.39 &  0.550 &   1.77  &    2.47  &       0.28 &  2.72 &  0.39    \\
HD-5004A &  3.94 &   3.26 &  0.67 &  0.840 &   1.60  &   2.20  &       0.37 &  4.00 &  0.43    \\
HD-5004B &  3.94  &   3.05 &  0.86 &  0.986 &   1.52  &    1.96  &       0.26 &  3.29 &  0.30    \\
HD-5103B &  3.78 &   2.63 &  0.46 &  0.820 &   1.92  &    3.07  &       0.55 &  3.90 &  0.72    \\
LD-5101A &  3.79 &   1.29 &  0.33 &  0.622 &   1.65  &    1.99  &       0.32 &  3.83 &  0.19    \\
\hline
\hline
\end{tabular}
\label{SimDiskData}
\end{minipage}
\end{table*}

\subsubsection{GASOLINE}

The {\tt GASOLINE} galaxies are cosmological zoom simulations with initial conditions
derived from the McMaster Unbiased Galaxy Simulations
\citep[MUGS,][]{Stinson:2010}.

When gas becomes cool ($T < 15000$ K) and dense ($n_{th} > 9.3$
cm$^{-3}$), it is converted to stars according to a
Kennincutt~\textendash~Schmidt-like law with the star formation
rate $\propto \rho^{1.5}$. Stars feed energy back into the
surrounding gas. Supernova feedback is implemented using the
blastwave formalism \citep{Stinson:2006} and deposits $10^{51}$
erg of energy into the surrounding medium at the end of the
stellar lifetime of every star more massive than 8\,M$_\odot$.
Energy feedback from massive stars prior to their explosion as
SNe has also been included \citep[][as part of the MaGICC project]{Stinson:2013}.
To mimic the weak coupling
of this energy to the surrounding gas \citep{Freyer:2006}, we
inject pure thermal energy feedback, which is highly
inefficient in these types of simulations \citep{Katz:1992,
Kay:2002}. We inject 10\% of the available energy during this
early stage of massive star evolution, but 90\% is rapidly
radiated away, making an effective coupling of the order of
1\%.

Ejected mass and metals  are  calculated based on the
Chabrier IMF \citep{Chabrier:2003} and  are  distributed to the nearest neighbor
gas particles using the smoothing kernel \citep{Stinson:2006}.
Literature yields for SNII \citep{Woosley:1995} and SNIa
\citep{Nomoto:1997} are used. Metals are diffused by treating
unresolved turbulent mixing as a shear-dependent diffusion term
\citep{Shen:2010}, allowing proximate gas particles to mix
their metals. Metal cooling is calculated based on the diffused
metals.

\subsection{Disk Galaxies}
\label{DiskGal}

\subsubsection{Simulated Disk Galaxies}
\label{SimuDisks}

In this case, simulations use the cosmological "zoom-in" technique, with
high-resolution gas and dark matter in the  region of the main
object. The cosmological parameters of a $\Lambda$CDM model
were assumed for P-DEVA ({\tt GASOLINE}) runs ($\Omega_{\Lambda}=0.723 (0.760)$, $\Omega_m=0.277 (0.240)$,
$\Omega_b=0.04 (0.04)$, and $h=0.70 (0.73)$), in a 10 Mpc (64 Mpc) per side periodic
box. 


\begin{figure*}
\centering
\includegraphics[width=.9\textwidth, angle=0]{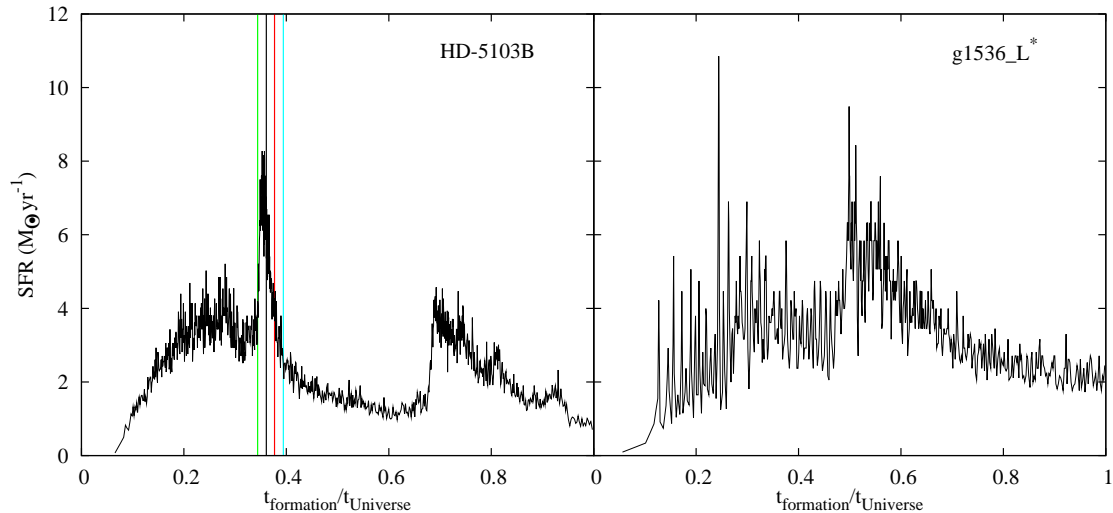}
\caption{Left-hand panel: the stellar age distribution at $z=0$ for the disk
galaxy HD-5103B from \citet{Domenech:2012}, where color vertical lines mark the snapshots
where this starbursting period will be analyzed in Section 4.  Right-hand panel:
the same for the g1536\_L$^{*}$ galaxy from MAGICC runs \citep{Brook:2012b}. }
\label{Age_d_z0}
\end{figure*}

As a first application, the SED of the HD-5103B galaxy,
analyzed by \cite{Domenech:2012}, has been calculated and
analyzed at $z=0$.
Moreover, the SEDs and colors of
HD-5103B around one of its star-forming major mergers have been
carefully studied and compared to the phase of milder SF activity at $z=0$.
To compare with currently available data on local, non-starbursting spiral galaxies,
see $\S$\ref{ObsData} below, we have also analyzed 4 more galaxies run with P-DEVA
\citep[LD-5003A, HD-5004A, HD-5004B, LD-5101A, see][]{Domenech:2012}, as well as 3 galaxies run  
with {\tt GASOLINE} from the MAGICC project \citep[g1536\_L$^{*}$, g21647, g7124, see][]{Brook:2012b,Stinson:2013}.
Except for g21647 and g7124 galaxies, first analyzed by Obreja et al.
(2014, submitted to MNRAS), these galaxies have all previously appeared
in the literature, where more details can be gathered. The
choices of most of the relevant parameters to run the P-DEVA
simulations, as well as many of those characterizing the
galaxies properties, are summarized in Tables 1 and 2 of
\citet{Domenech:2012} and in Table 1 of \citet{Obreja:2013}.

An important parameter here is the softening, setting  the
grid cell size. The gravitational softening used in the 5 P-DEVA runs
is $\epsilon_g = 400 h^{-1}$ pc, while $\epsilon_g = 312,5$ pc for g1536\_L$^{*}$ and g7124,
and $\epsilon_g = 156,2$ pc for g21647 galaxies. 
Some  properties of the $z=0$ galaxies,
provided by the simulations and some of them relevant to GRASIL-3D SED calculation,
are given in Table~\ref{SimDiskData}. We see that both the SFR (averaged over the past 100 Myrs) and the specific SFR
are low, that the disk and bulge scalelengths, and the bulge-to-disk luminosity ratios are consistent with
observations, as  other galaxy properties previously analyzed \citep[see, for example][]{Domenech:2012,Stinson:2013}.
The images of these galaxies \citep[see for example Figure~\ref{DiskImage} for g1536\_L$^{*}$ or Figure 1 in][]{Domenech:2012}  show that they are not interacting galaxies,
as defined for example by \citet{Smith:2007} or \citet{Lanz:2013}, see $\S$~\ref{ObsData} below.
Therefore, these 8 galaxies can be considered as local normal galaxies,
with low or very low starbursting activity, and hereafter will be refereed to as the
$z=0$ sample of normal simulated disk-like galaxies.

More details on this $z=0$ sample of simulated disk-like galaxies provided by GRASIL-3D
can be found in Table~\ref{MassLumDisk}, and on 
HD-5103B  during its starbursting phase in
Table~\ref{MCresults}. Results in both Tables will be discussed later on.

To calculate the SED of a galaxy, a crucial piece of
information is its stellar age distribution. They are provided
in Figure~\ref{Age_d_z0} for HD-5103B and g1536\_L$^{*}$, where we can see that HD-5103B has
been involved in two  major merger events during its assembly
process, with starbursting activity.
The first starbursting phase will be analyzed in detail in  $\S$\ref{DiskResults}
as a model for a system with relatively high SF activity.
\footnote{
This particular starburst phase has been chosen
because it is the strongest one in our set of simulated spiral
galaxies (see Dom\'enech et al. 2012), however the results are
shown  transformed to $z$s typical of those of local galaxies.
}.  

Note that to transform from stellar age distribution to star formation rate,
needed to calculate  stellar luminosities, we corrected for gas restitution
applying the \citet{Lia:2002} formalism \citep[see also][]{Martinez:2008}.


\begin{figure}
\includegraphics[width=.5\textwidth, angle=0]{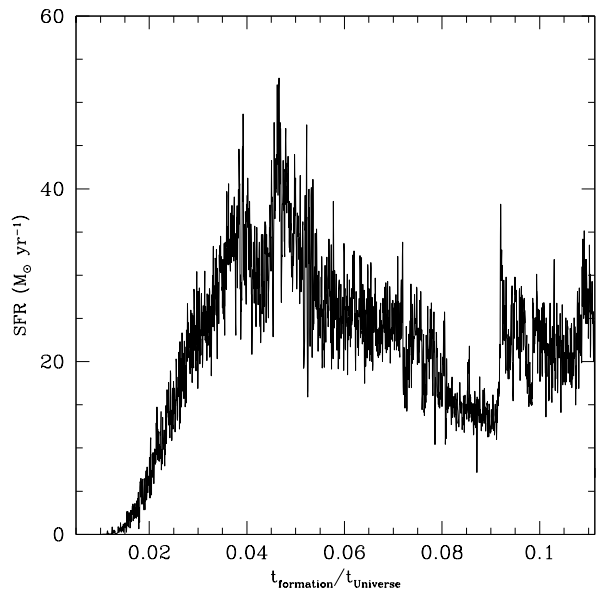}
\caption{The stellar age distribution at $t_{\rm formation}/t_{\rm Universe}=0.1113$ of the high $z$ galaxy, D-6254, identified in the simulation \#7629.} 
\label{sfr7629}
\end{figure}


\subsubsection{The Parameter Space and its Validation}
\label{ParSpace}

The allowed ranges of parameter values in GRASIL-3D have been
discussed in Section~\ref{Parameters}. 
More specifically, in Table~\ref{DiskParTable}
we give the combinations of parameters we used for studying the $z=0$
sample of normal simulated disk-like galaxies (those marked N in column 2),  as well as the starbursting phases of HD-5103B
(marked S in column 2).
Each
parameter set is identified by its name (first
column) and a symbol  in Figure~\ref{FIRoB}. 


\begin{table}
\centering \caption{The parameter combinations used to test GRASIL-3D on simulated disk galaxies, either in starbursting phases (S) or  in  normal (N) ones}
\begin{tabular}{c cccc}
 \hline
Set  &  Phase & $t_0$ &  $\rho_{mc, thres}$ & $\sigma$     \\
     &   & (Myrs)&  (M$_{\odot}$kp$^{-3}$) &                 \\
\hline
\hline
& & $r_{mc}=$14 pc & &  \\
\hline
\hline
 1  & N & 2.5      &      3.3$\times 10^9$          &  3         \\ 
 2  & N & 2.5      &      3.3$\times 10^9$          &  2          \\ 
 3  & N & 2.5      &      3.3$\times 10^8$          &  3          \\ 

\hline
 4  & N & 5      &      3.3$\times 10^9$          &  3           \\ 
 5  & N & 5      &      3.3$\times 10^9$          &  2          \\ 
 6  & N & 5      &      3.3$\times 10^8$          &  3           \\ 
\hline
 7 & N, S & 10      &      3.3$\times 10^9$          &  3      \\ 
 8  & N, S & 10      &      3.3$\times 10^9$          &  2      \\ 
 9  & N, S & 10      &      3.3$\times 10^8$          &  3      \\ 
\hline
  10  & S & 40      &      3.3$\times 10^9$          &  3     \\ 
 11  & S & 40      &      3.3$\times 10^9$          &  2     \\ 
 12 & S & 40      &      3.3$\times 10^8$          &  3     \\ 
\hline
\hline
& & $r_{mc}=$17 pc & &  \\
\hline
\hline
 13  & N, S & 10      &      3.3$\times 10^9$          &  3       \\ 
 14  & N, S & 10      &      3.3$\times 10^9$          &  2       \\ 
 15  & N, S & 10      &      3.3$\times 10^8$          &  3       \\
\hline
\hline
\end{tabular}
\label{DiskParTable}
\end{table}


To make  identification easier, the
table is divided by a double horizontal line, and each part is
further divided by simple horizontal lines. The  difference
between  the sets belonging to the upper and lower Table blocks
are the parameters characterizing individual MCs, namely their
radii $r_{mc}$, entering in the calculation of their optical
depth, now a local quantity (see Eqs. \ref{taumc} and
\ref{dustfrac}). Parameter sets separated by single horizontal
lines have  different $t_0$ values (we recall that this parameter
sets the fraction of light that can escape the starburst
regions, mimicking MC destruction by young stars, see
Section~\ref{yslum}). 

Finally, the three parameter sets within
each sub-block differ in the parameters setting the molecular
mass (and cirrus) fraction at each grid cell, namely the
threshold density for MC formation and the dispersion in the
PDF, $\rho_{mc, thres}$ and $\sigma$ respectively. They also
set the global MC and cirrus mass, as well as the global MC
fraction, $f_{mc}$, and through the cirrus mass, the total
amount of dust in cirrus.


\begin{figure*}
\centering
\includegraphics[width=1.0\textwidth, angle=0]{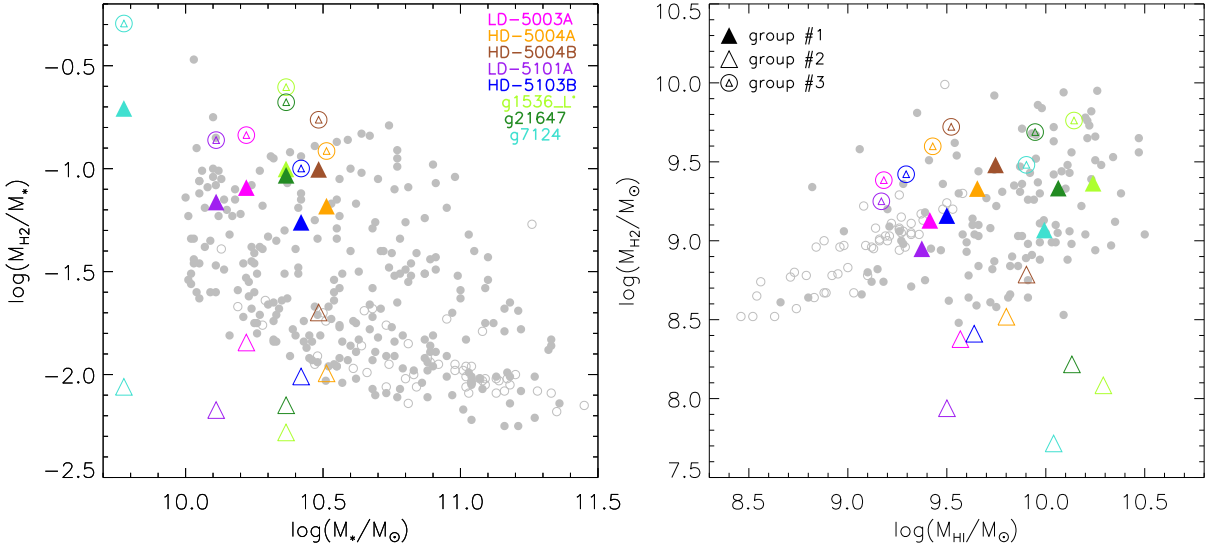}
\caption{Comparison of molecular and atomic hydrogen content and stellar masses of the simulated sample of disk galaxies 
 at $z=0$    to those of galaxies in the COLD GASS survey
(Saintonge et al. 2011). Different color symbols distinguish different objects 
according to the coding in the left-hand panel. Filled, open and composed symbols correspond to parameter Sets \# (1,4,7,13), (2,5,8,14) and (3,6,9,15),
respectively, these three groups giving degenerated results in these plots.  Gray points  are data, with open
symbols corresponding to non-detections (upper limits). For details see Saintonge et al.
(2012)  and Figures 5 and  8  therein. } \label{HI_H2}
\end{figure*}


The models in Table~\ref{DiskParTable} explore the entire range
of allowed values for the $t_0$ parameter ($t_0 \sim$ 2.5 - 8 Myrs for normal spiral galaxies,
and $t_0 \sim 18 - 50$ Myrs for starbursting ones, according to S98), as well as the range
for the $\rho_{mc, thres}$ and $\sigma$ parameters. Regarding these last two:
 i), the
extreme values of $\rho_{mc, thres}$ are  10 - 100 H nuclei
cm$^{-3} \approx 3.3 \times 10^9 - 3.3 \times 10^8 M_{\odot} \times kpc^{-3}$
taking a H nuclei-to-gas mass ratio of 0.75, and
 ii), for the
dispersion of the log-normal PDF we have also taken its extreme
values found in literature, $\sigma = 3$ and $\sigma = 2$. This
would give 4 combinations describing extreme range values for
the parameters governing the calculation of $f_{mc}$. It turns
out that the combinations ($\rho_{mc, thres} = 3.3 \times 10^9 M_{\odot}kpc^{-3}, \sigma
= 3$) and ($\rho_{mc, thres} = 3.3 \times 10^8 M_{\odot}kpc^{-3}, \sigma = 2$) give
similar results, and therefore we have not shown the last.

As discussed in $\S$\ref{DustCompPro},
only the    $\rho_{mc, thres} $ and $\sigma$ parameters affect the MC content
of a given galaxy. Indeed, in Table~\ref{MCresults} the degeneracies
among parameter Sets relative to masses of molecular and atomic
hydrogen of simulated galaxies are clearly shown.
 In fact, using  parameter Sets belonging to each of the following three groups 
 \# (1,4,7,13), \# (2,5,8,14) and \# (3,6,9,15), 
 gives  the same MC and cirrus content. Hereafter they will be named groups \#1, \#2 and \#3 of parameter Sets, respectively. 

An important and necessary consistency test for these choices
is to compare the resulting masses of molecular and atomic
hydrogen of simulated galaxies to observational data. This has
been done in Figure~\ref{HI_H2} for local galaxies, where gray
symbols are data on local galaxies taken from
\citet{Saintonge:2012}, and color symbols correspond to the sample of 8  $z=0$  simulated disk-like
galaxies described above, see also Table~\ref{MassLumDisk} for parameter Set~4.
More information on the MC content of
some of the simulated galaxies can be found in Table~\ref{MCresults}
for the different parameter Sets,
where we see that the mass in their MC content increases 
when $\sigma$ increases, from
group \#2 of parameter Sets to group \#1, and  increases more again when $\rho_{mc, thres}$ is reduced from
  3.3 $\times 10^9$ M$_{\odot}$kp$^{-3}$  in group \#1 to  3.3  $\times 10^8$ M$_{\odot}$kp$^{-3}$ in group \#3. 

Then, in
Figure~\ref{HI_H2}, we see that group \#2 of parameter Sets  results
in too little molecular gas  
   for the three galaxies run with {\tt GASOLINE}
(open green and turquoise triangles
outside the data cloud), as well as for the LD-5101A  one run with P-DEVA.
Therefore group \#2 of parameter Sets will not be further used in this paper
for these galaxies.
However, the extreme $f_{mc}$ values
provided by group \#2 and group \#3 are within the
range of observations for   the other galaxies. Overall,   of the 3 models in
Figure~\ref{HI_H2} we see that
 the parameter Sets within group \#1  give the best (and
comfortably good) consistency with \citet{Saintonge:2012} data
of local galaxies. We also note that the stellar mass of
g7124 is outside the rage of stellar masses in \citet{Saintonge:2012}.

As remarked in $\S$\ref{Parameters}, PHIBSS survey on $z \sim
1-3 $ massive galaxies ($M_{*} \ge 10^{10.4}$) by
\citet{Tacconi:2013}, provides new data on their molecular
content. An extrapolation to lower masses is provided by their
Figure 12, to be compared to our results in
Table~\ref{MCresults} for the merging phase of HD-5103B. The
consistency is good for starburst galaxies\footnote{We note
that starburst galaxies are likely to need a different
conversion factor from the observed CO line flux to molecular
gas mass ($\alpha \sim 1$ instead of 4.36).}. We conclude that the values used for $\rho_{mc,thres}$  and
$\sigma$  adequately describe the molecular cloud content
of simulated disk galaxies.

Regarding other parameters, we note that the values of both the
time young stars remain enshrouded  in MCs ($t_0=2.5$ Myrs, parameter Sets 1 - 3,
$t_0=5$ Myrs, parameter Sets 4 - 6, and 10
Myrs, parameter Sets 7 - 9), and the masses and radii of molecular clouds ($m_{mc}=
10^6$M$_{\odot}$ and $r_{mc}= 14$pc)  we have used are typical
of normal spiral galaxies according to the discussion in
Section~\ref{Parameters}. We have also used a value of $t_0=40$ Myrs, 
more typical of starburst galaxies, to
test the star-forming phases of these galaxies (parameter Sets 10,
11 and 12). Values of $t_0=10$ Myrs in combination with a molecular cloud radius $r_{mc}= 17$ pc (parameter Sets
 13, 14 and 15) have been used   to find out the effects of having a
smaller optical depth in MCs (see Eq. (31)). 
The effects of
GRASIL-3D parameter variations on the SEDs of simulated
galaxies will be further discussed in Section~\ref{ParDiscu}.


\begin{figure*}
\centering
\includegraphics[width=.37\textwidth, angle=270]{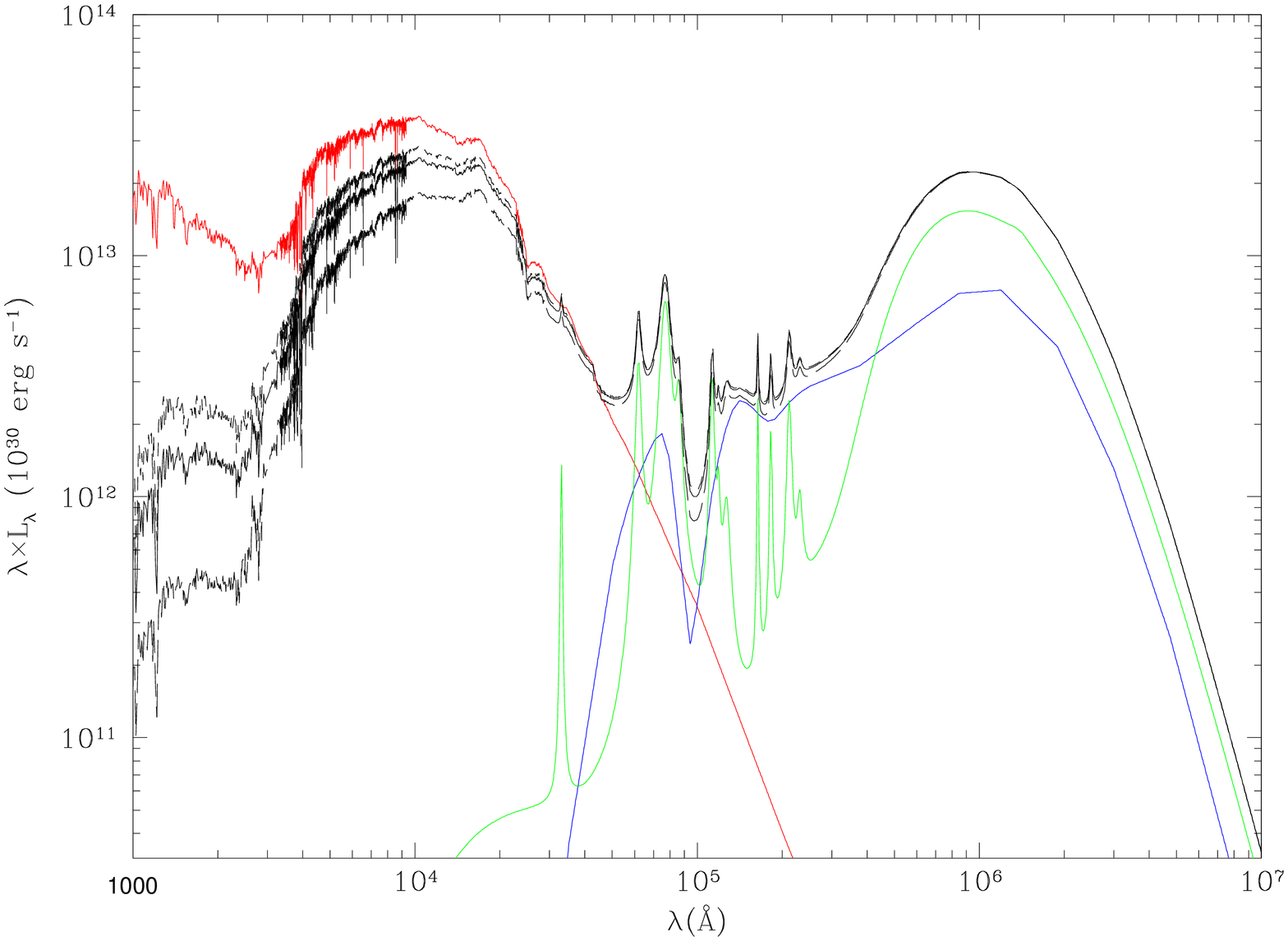}
\includegraphics[width=.37\textwidth, angle=270]{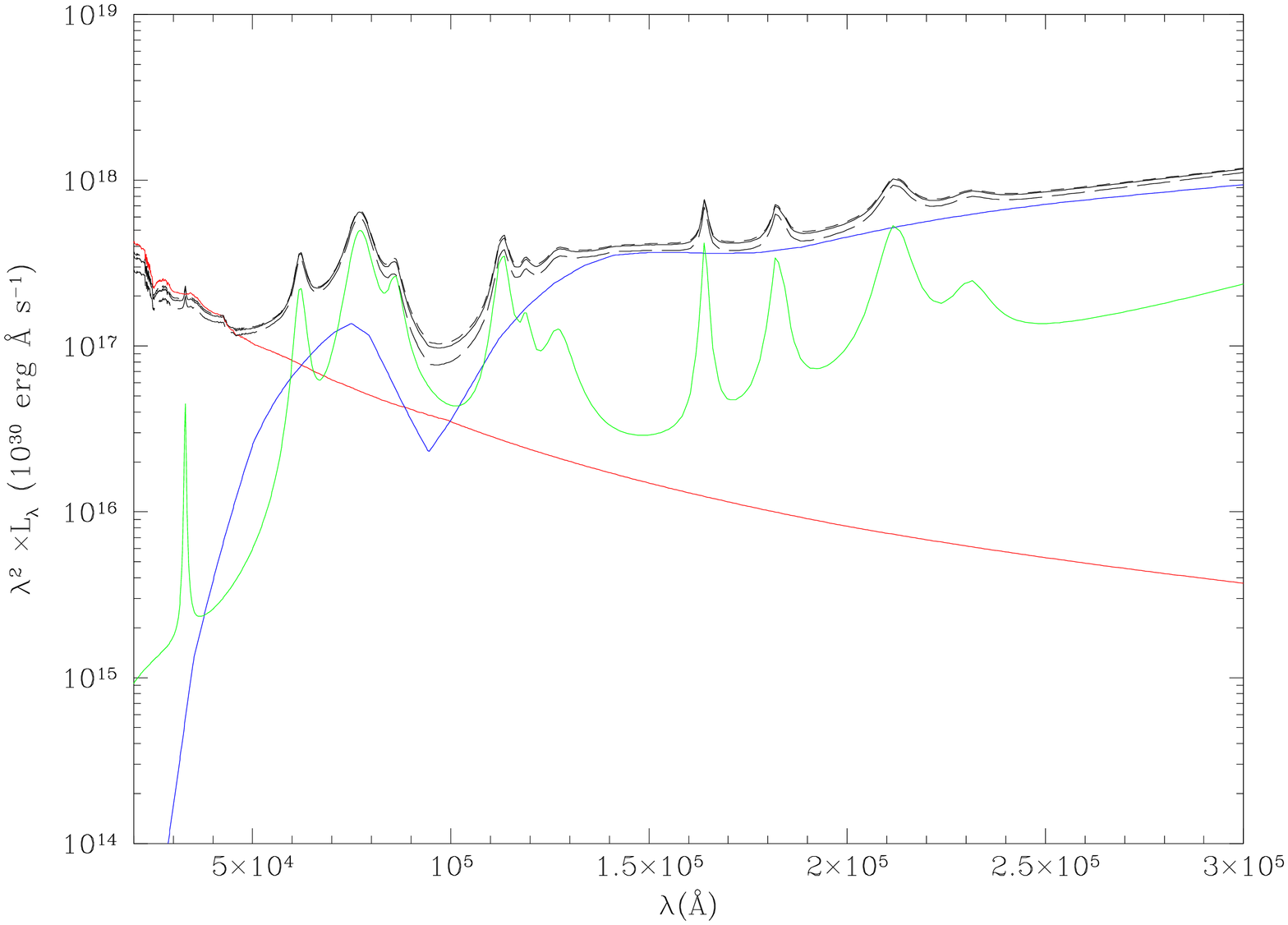}\\
\includegraphics[width=.37\textwidth, angle=270]{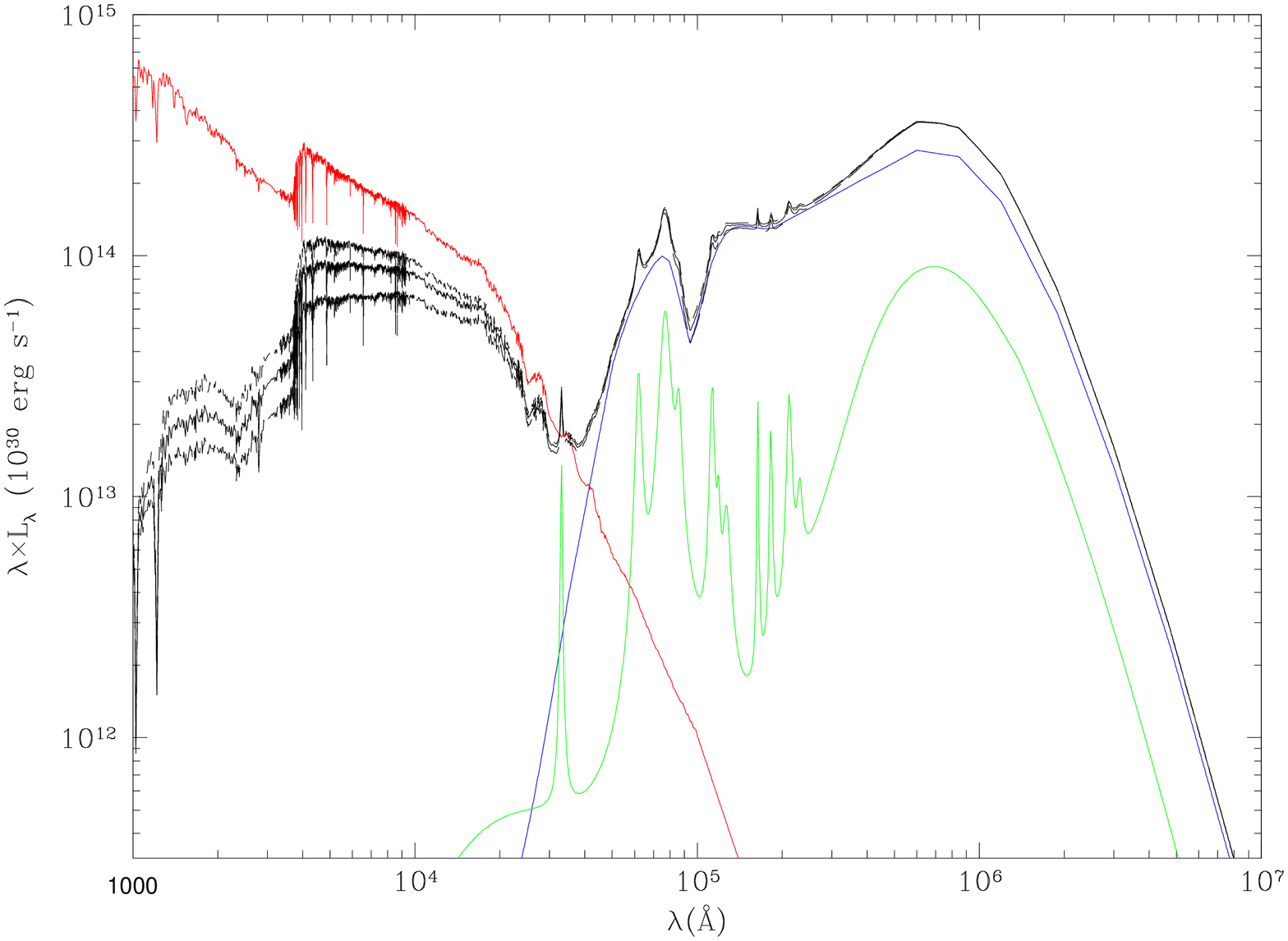}
\includegraphics[width=.37\textwidth, angle=270]{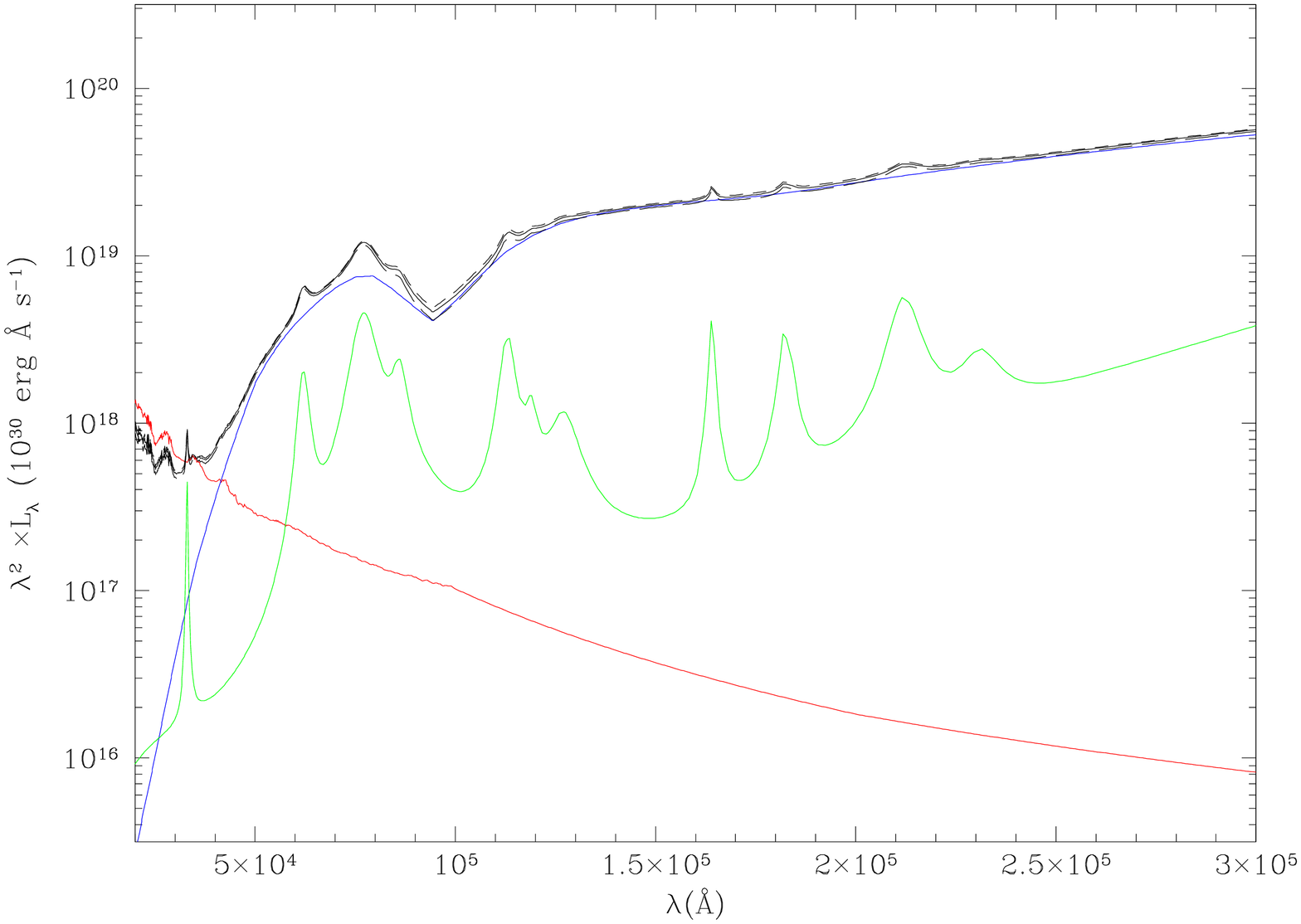}
\caption{The rest-frame SEDs of disk galaxy HD-5103B at $z=0$ calculated with  parameter Set~4 (top left) and of starburst D-6254 with parameter set SB7 (bottom left), and
the zooms  in the PAH band region of the rest-frame emitted fluxes of HD-5103B (top right) and D-6254 (bottom right).
The color code for the lines is the same as in Figure 1, with the upper (lower) black line corresponds to the face-on (edge-on) view of the galaxies,
while the  middle black line is the angle-averaged emission.}
 \label{Disk_51.50}
\end{figure*}


\begin{table*}
\begin{minipage}{6.5in}
\renewcommand{\thefootnote}{\thempfootnote}
\centering \caption{Masses, luminosities and energy balances (in parentheses) for the sample of $z=0$ disk-like galaxies calculated with parameter Set \# 4 }
\begin{tabular}{l ccccccc}
\hline
Name & Young Stars & Free Stars  & MCs & Dust in Cirrus &  $L_{bol}$ &  $L_{c}$ & $L_{mc}$ \\
& ($10^{6} M_{\odot}$) & ($10^{10} M_{\odot}$)   & ($10^{10} M_{\odot}$)   & ($10^{7} M_{\odot}$)   & ($10^{44}$ erg sec$^{-1}$) & ($10^{44}$ erg sec$^{-1}$) & ($10^{44}$ erg sec$^{-1}$)  \\
\hline
\hline
g1536\_L$^{*}$   & 18.269 & 2.3188  & 0.2312  & 9.5034  & 1.7857 (97.9)  & 0.4948 (98.4) &  0.5506 (92.2) \\
g21647  & 46.906 & 2.3171  & 0.2155  & 9.5589  & 3.2408 (99.4)  & 1.2932 (96.9) &  1.2597 (96.5)  \\
g7124     & 7.4456 & 0.5959  & 0.1169  & 3.8345  & 0.5922 (94.6)  & 0.0715 (99.3) &  0.2448 (87.1)           \\
LD-5003A  & 6.4977 & 1.6624  & 0.1346  & 2.8819  & 0.4676 (99.9)  & 0.1197 (98.2) &  0.1019 (94.2) \\
HD-5004A  & 8.2845 & 3.2545  & 0.2140  & 4.2582  & 0.8334 (99.2)  & 0.2413 (98.4) &  0.1629 (95.0) \\
HD-5004B  & 10.651 & 3.0469  & 0.3023  & 4.5436  & 0.8426 (99.6)  & 0.2561 (99.4) &  0.1761 (94.2) \\
HD-5103B  & 4.9310 & 1.2906  & 0.0889  & 2.1182  & 0.4090 (100)   & 0.0739 (96.7) &  0.0681 (93.6) \\
LD-5101A  & 6.0475 & 2.5362  & 0.1433  & 4.0385  & 0.7864 (98.8)  & 0.2303 (98.7) &  0.1471 (97.8) \\
\hline
\hline
\end{tabular}
\label{MassLumDisk}
\end{minipage}
\end{table*}


\begin{figure*}
\includegraphics[width=1.0\textwidth,  angle=0]{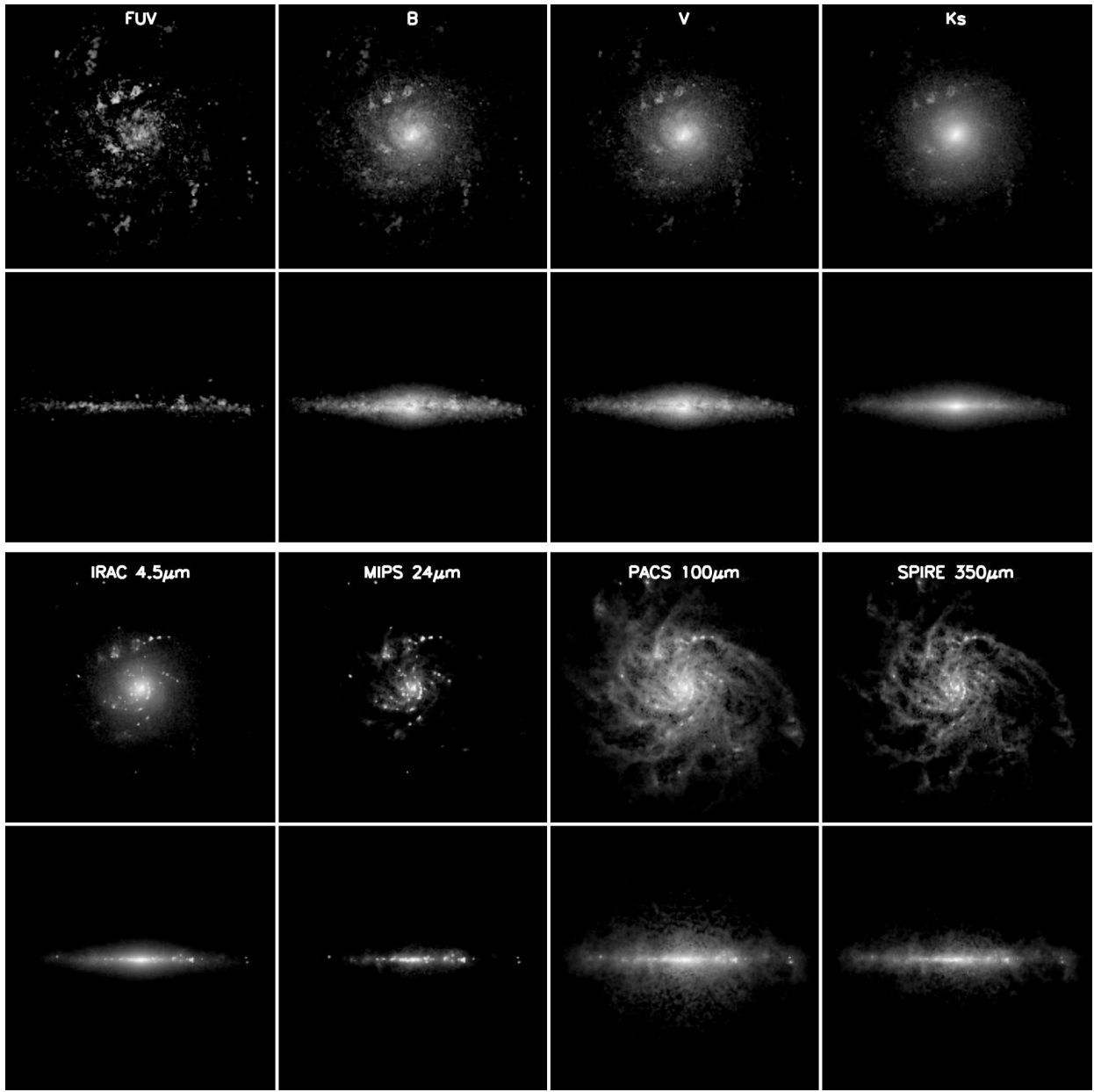}  
\caption{Face-on and edge-on images of the g1536\_L$^{*}$ galaxy at redshift
$z=0$ in the 8 bands specfied in each panel. 
  The  physical size of each panel is
50 kpc per side.
No telescope effects (like point-spread functions, pixel sizes, etc) have been taken into account. }
\label{DiskImage}
\end{figure*}
\begin{figure*}
\includegraphics[width=1.0\textwidth,  angle=0]{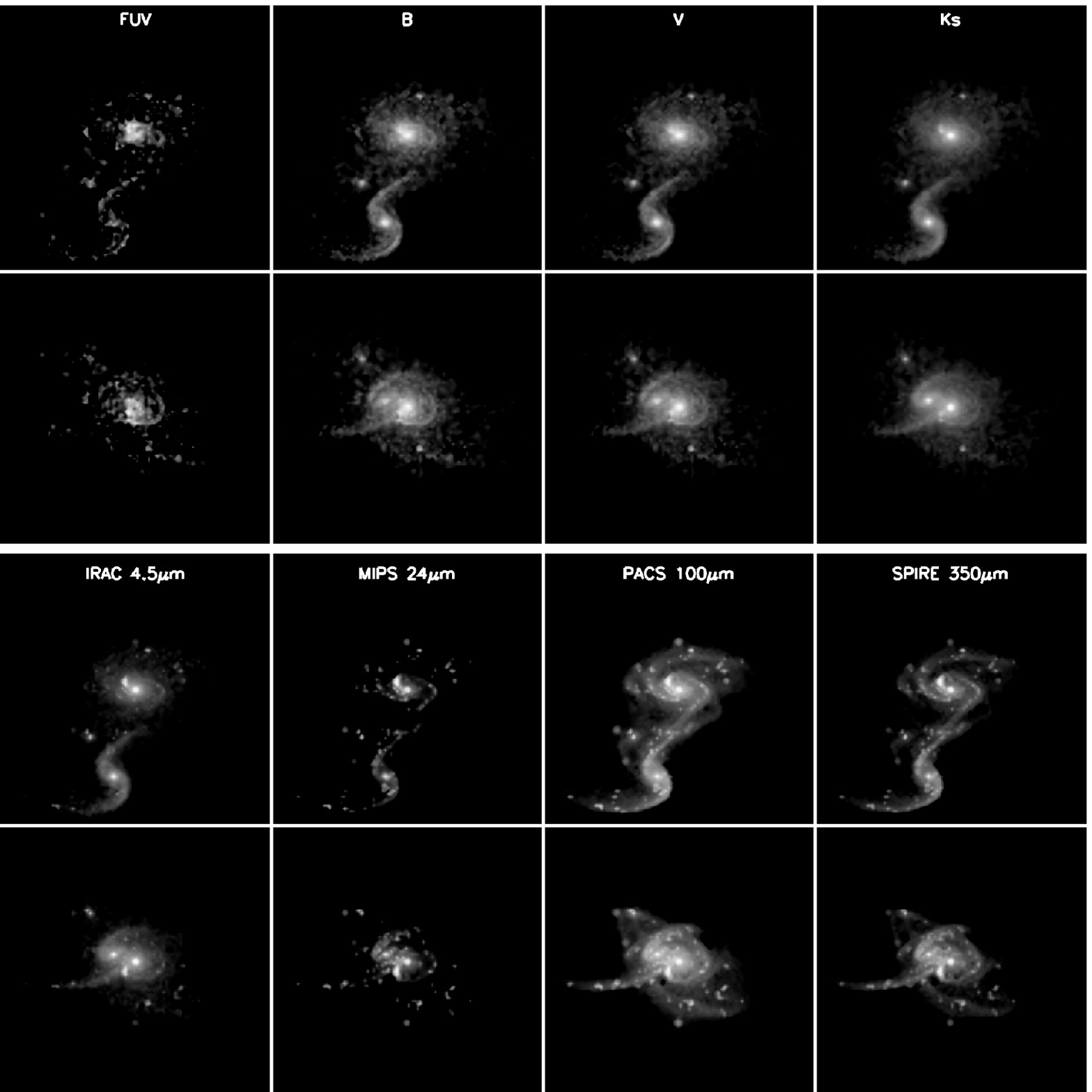}  
\caption{Two orthogonal views of a triple merger at $z=0.28$ involving disk galaxies giving rise to  HD-5103B spiral at $z=0$.
The 8 bands specfied in each panel are the same of the previous Figure.  The  physical size of each panel is
45 kpc per side.
No telescope effects (like point-spread functions, pixel sizes, etc) have been taken into account.}
\label{MergerImage}
\end{figure*}

\subsubsection{The SEDs and Luminosities  of Simulated Disk Galaxies}
\label{SedSimuDisks}

To demostrate how the SEDs of simulated galaxies 
look, in
Figure~\ref{Disk_51.50} we show the SED of HD-5103B disk galaxy
(top left) and a zoom of the corresponding emitted flux 
in the MIR region (top right), where the PAH band emission stands out. 
This SED has been calculated using the fiducial
dust model described  in Section 2.4 and the 
parameter Set~4 given in Table~\ref{DiskParTable}. 
Their color code is similar to that in Figure 1, with
the upper (lower) black lines corresponding to the face-on (edge-on)
view of the galaxies, while the  middle black line is the
angle-averaged emission. We can see in Figure~\ref{Disk_51.50} that, as expected, 
the difference  between the face-on and edge-on views is remarkable
in the UV and at optical wavelengths, while it is unimportant at longer ones.
The comparison of these SEDs to observations will be analyzed in $\S$\ref{CompaObs}.

Other important data on simulated galaxies returned by GRASIL-3D, appart from those
shown in Figure~\ref{HI_H2}, are the total bolometric luminosity, $L_{bol}$, and the total cirrus
and molecular cloud emmitted energy, $L_{c}$ and $L_{mc}$, respectively. These can be found
in Table~\ref{MassLumDisk}    for the $z=0$ sample of normal simulated 
disk-like galaxies, along with their respective energy balances. We see that the luminosities are
consistent with observations (see  $\S$\ref{CompaObs} below), and that
the balances are generally  very good.

\subsection{Simulated Disk and Merger Images}
\label{ImagesResults}

 In Figure~\ref{DiskImage}  we show face-on and edge-on images of the g1536\_L$^{*}$ galaxy at redshift
$z=0$ in 8 bands going from far-UV to FIR. From left to right and 
from top to bottom, they  correspond to GALEX (FUV), Johnson (B and V), 2MASS (Ks), IRAC (4.5 $\mu$m), MIPS (24 $\mu$m),
PACS (100 $\mu$m) and SPIRE (350 $\mu$m) bands.
The  physical size of each panel is
50 kpc per side.
Note  the clumpy appearence in the UV and IR bands, as compared with the other bands.   The effects of dust obscuration are clear in the edge-on images, and particularly so in the B and V bands where dust lanes can be appreciated accross the disk.
 Bright spots are star formation regions, nicely visible on the spiral arms.

  GRASIL-3D is very useful to study   mergers of simulated galaxies, where geometries can be very complex. As an illustration, in Figure~\ref{MergerImage}  we show two orthogonal views of a triple merger of disk galaxies at $z=0.28$,  giving rise to the HD-5103B spiral at $z=0$. The bands are the same as in the previous Figure (rest-frame emission). In this case the panels correspond to 45 kpc per side. Again, the UV and IR emission are more clumpy than the other bands. Note that in one of the perspectives, tidal tails (where IR emission concentrated in knots corresponds to star formation bursts) look very similar to those observed in the local Antennae merger, while in the other the spiral structure of one of the galaxies is still visible.  


\subsection{Comparing GRASIL-3D SEDs of Disk Galaxies to Observations}
\label{CompaObs}

It is important to know how well the SEDs of simulated galaxies
behave as compared to observational data. We are particularly interested in the rest-frame near-IR region up to the
far-IR, where the effects of cirrus and MC emission dominate.
Comparisons with data coming from different projects are
discussed in turn, taking into consideration the starbursting/quiescent SF  phase and/or the interacting/non-interacting situation of simulated and real galaxies.

\subsubsection{Observational Data to Compare with}
\label{ObsData}


\begin{figure*}
\centering
\includegraphics[width=1.0\textwidth, angle=0]{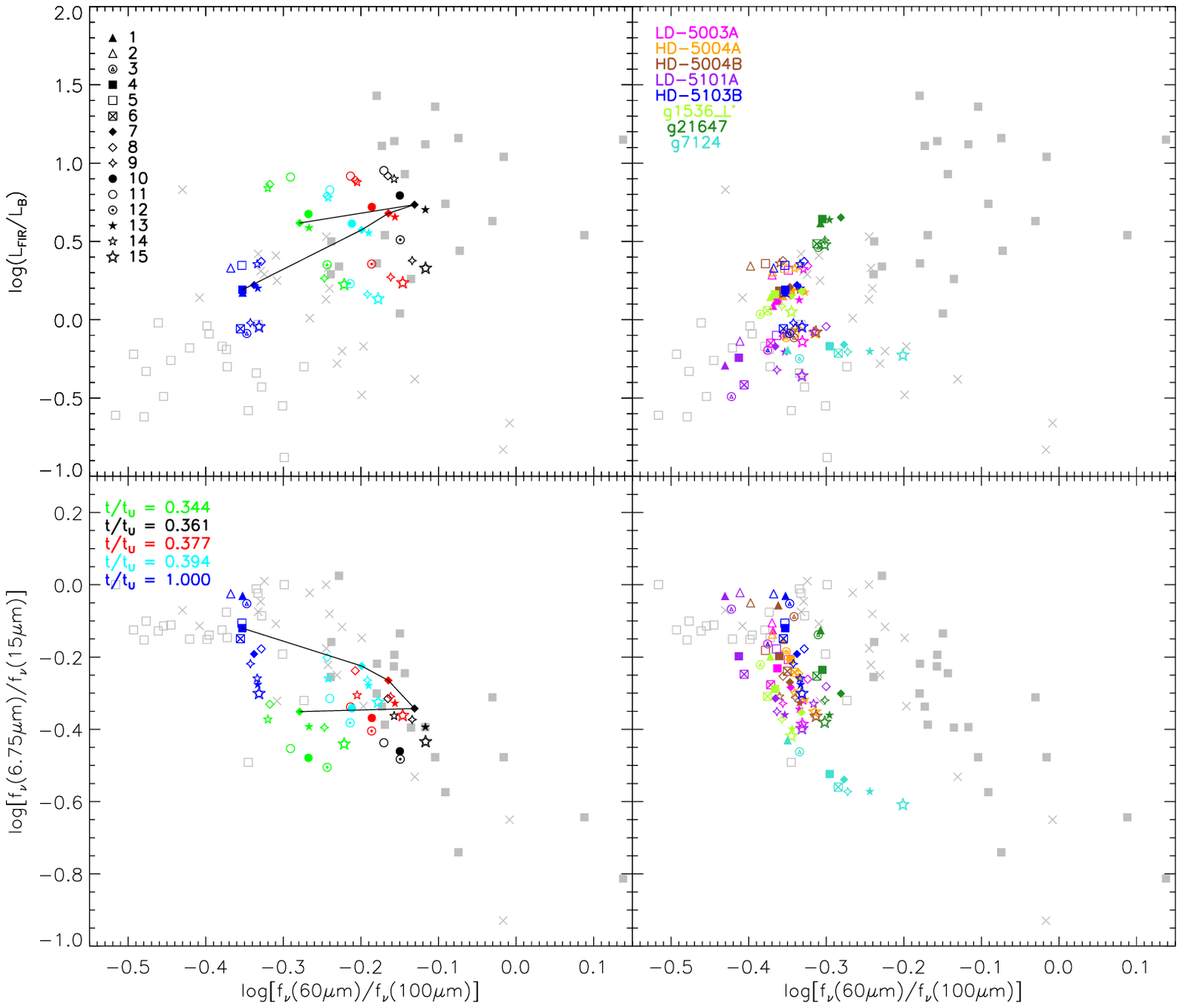} 
\caption{Upper-left panel: The rest-frame FIR/blue luminosity ratio $L_{\rm FIR}/L_{\rm B}$
versus the rest-frame IRAS flux density ratios f$_{\nu}$(60$\mu$m)/f$_{\nu}$(100$\mu$m)
for the HD-5103B galaxy at $z=0$ (blue) and at different times around one of its major mergers,  according to the color code in the lower left-hand panel. Different symbol shapes distinguish different parameter sets
(see Table~\ref{DiskParTable}) as encoded on the left of this panel. Gray points are the same
ratios for the \citealt{Dale:2000} sample shown for comparison, where the filled and
empty squares represent FIR-active (i.e., starbursting galaxies) and FIR-quiescent objects, while crosses are intermediate
systems (mild starbursts), according to the classification of \citealt{Vega:2005}.
Lower-left panel: Same as in the upper panel for the f$_{\nu}$(6.7$\mu$m)/f$_{\nu}$(15$\mu$m) vs the f$_{\nu}$(60$\mu$m)/f$_{\nu}$(100$\mu$m) flux ratios. 
Black lines in these two panels join consecutive parameter Set~7 results for the HD-5103B galaxy
at its starburst phases and the Set~4 result for the quiescent phase at $z=0$.
The right-hand panels are similar to those at their left for the whole sample of 8 $z=0$ simulated normal disk-like galaxies, whose identity is encoded by the colors at the left of this panel. 
Face-on  emissions have been used to draw this Figure.
}

\label{FIRoB}
\end{figure*}

The following observational samples and projects are used as reference:
\begin{enumerate}
\item ISO Key Project on the ISM of Normal Galaxies \citep{Helou:1996}. \citet{Dale:2000} provide ISO and IRAS
broad-band fluxes. Based on the \citet{Lu:2003} definition of stellar activity, V05 classify these galaxies into FIR-active (i.e., starbursting, with log[L$_{\rm FIR}$/L$_{\rm B}$] $\ge 0$ and
log[f$_{\nu}$(60$\mu$m)/f$_{\nu}$(100$\mu$m)] $ \ge -0.24$, 43\% of the sample), FIR-quiescent (with no or very low SF, 
log[L$_{\rm FIR}$/L$_{\rm B}$] $< 0$ and 
log[f$_{\nu}$(60$\mu$m)/f$_{\nu}$(100$\mu$m)] $  < -0.24$, 31\% of the sample) and FIR-intermediate (mild SB phases, 26\% of the sample), see their Figure 3.

\item \citet{Lu:2003} sample: a subsample of \citet{Dale:2000} where
the aromatic features in emission
(hereafter AFEs) at 6.2, 7.7, 8.6, and 11.3 $\mu$m dominating
the mid-infrared (MIR) are analyzed in detail. In addition, Lu et al. provide their  ISOPHOT spectra.

\item The Spitzer IR Normal Galaxy Survey (SINGS), see \citet{Kennicutt:2003} and \citet{Dale:2005}: 75 local
galaxies with different morphologies with Spitzer data.
By removing SINGS galaxies with close companions, \citet{Smith:2007} have selected a sample of 42  non-tidally perturbed galaxies (their non-interacting sample; 26 spirals, 4 E+SO, 12 Irr/Sm galaxies). 
A subsample of these galaxies using a more conservative definition for being non-interactive  is listed in Table 10 of \citet{Lanz:2013}  (as part of the KINGFISH project). 

\item The Herschel project on Key Insights on Nearby Galaxies: Far Infrared Survey with Herschel \citep[KINGFISH,][]{Kennicutt:2011}, which consists of 61 galaxies where 57 are SINGS galaxies. 
They are subluminous IR galaxies and all normal galaxy types are represented. All were imaged with Herschel PACS and SPIRE \citep{Dale:2012}, including those
listed by \citet{Lanz:2013} as non-interactive. 

\end{enumerate}

\subsubsection{Results}
\label{DiskResults}


\begin{figure}
\centering
\includegraphics[width=.48\textwidth, angle=0]{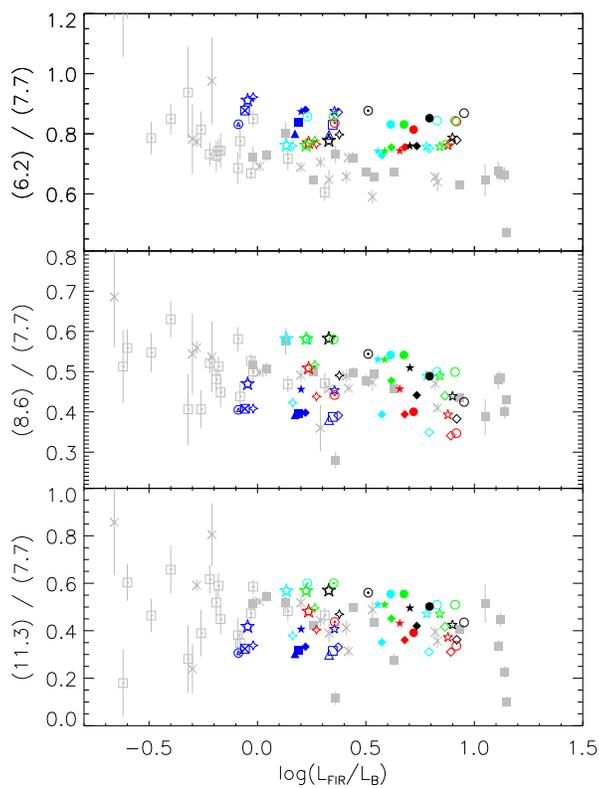}
\caption{The relative strengths of the AFEs as a function of the FIR/blue luminosity ratio  for the HD-5103B galaxy at $z=0$ (blue) and at different times around one of its major mergres. 
The color and shape codes are the same as in Figure~\ref{FIRoB} (left-hand panels), with the gray points (filled and empty squares, and crosses)
representing the observational sample of \citealt{Lu:2003}, according to their nature (FIR-active, FIR-quiescent and intermediate, respectively).
}
\label{AFEratio}
\end{figure}


\begin{figure}
\includegraphics[width=0.49\textwidth, angle=0]{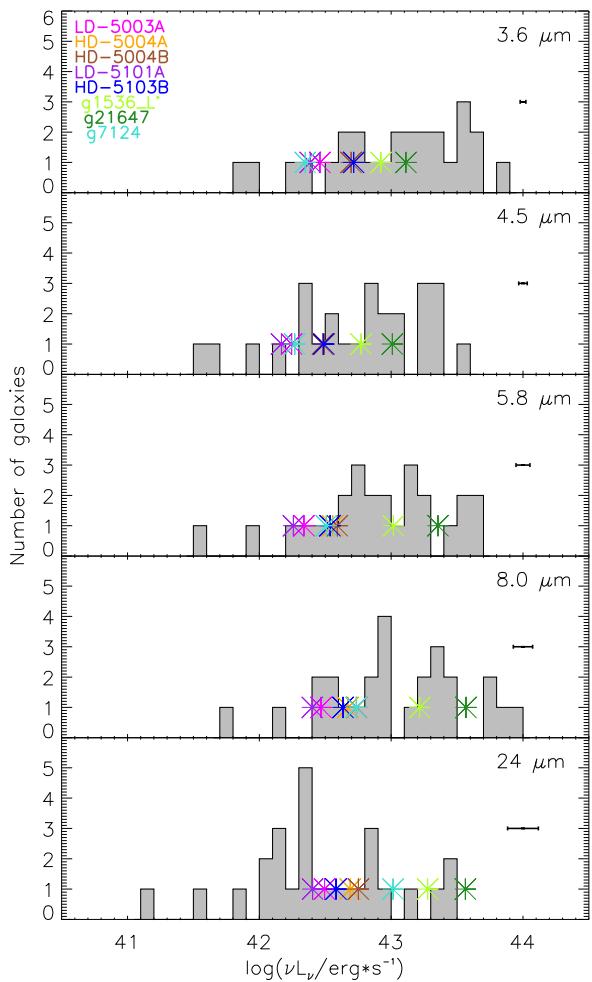}
\caption{Comparison of fluxes in different  Spitzer IRAC and MIPS bands for non-interacting normal disk-like galaxies.
The histograms have been drawn with data for the sample of 26 non-interacting galaxies from 
\citet{Smith:2007}, Tables 2 and 6, selected from SINGS galaxies, see text. Points stand for the sample of 8 disk-like simulated $z=0$ galaxies, their identity is encoded by colors as in previous Figures. Plotted are averages over results as model parameters take their allowed values (Table 2 for N phase). The error bars  at the right of each panel give the maximum dispersions over the 8 simulated galaxies. } 

\label{IRAC_fluxratio}
\end{figure}

As most of these projects provide us just with photometric data (i.e., no spectra),
 rather than a direct comparison of SEDs, an
appropriate method to compare calculated SEDs with observed
ones comes from fluxes, colors or flux ratio comparisons. As a first
exercise, 
we have made comparison to IRAS and ISOCAM results
on normal local galaxies, i.e., the Dale et al. 2000 sample,
classified by  V05 according to their SFR activity, see (i) in $\S$\ref{ObsData} above. 
In Figure~\ref{FIRoB} we plot the FIR/blue luminosity ratio $L_{\rm FIR}/L_{\rm B}$ versus the IRAS 60
$\mu$m/ 100 $\mu$m flux density ratio, for the Dale sample (in gray) with different symbols standing for
starbursting/FIR-active, mildy starbursting/FIR-intermediate and FIR-quiescent galaxies.

To test the ability of GRASIL-3D to return observational results for starbursting galaxies, the HD-5103B starbust period around $t/t_U \sim 0.35$ has been analyzed in detail 
by plotting its different phases in Figure~\ref{FIRoB}, upper panel at the left, and comparing them  with the situation at $z=0$, where the SF activity is milder. 
The different colors distinguish different starburst phases. More specifically, green, black, red, and cyan correspond to $t/t_U$ = 0.344, 0.361, 0.377, and 0.394, respectively. 
In Figure~\ref{Age_d_z0} we see that they correspond to the beginning of the starburst phase, two snapshots along the active phase (in black and red), and its end, respectively.
We also plot these ratios for the HD-5103B at $z=0$ (blue). Different symbols in the simulated galaxies stand for different parameter Sets, according to the code in the panel and to Table~\ref{DiskParTable}. 
Results for the different models in this Table have been drawn in  Figure~\ref{FIRoB} to illustrate their dispersion due to parameter variation 
within their allowed ranges (note that  parameter Sets with $t_0 = 40$ Myrs are excluded for $z=0$ galaxies). To highlight the time evolution, in Figure~\ref{FIRoB} 
those points corresponding to parameter Set \#7  (\#4) in the starburst (quiescent) phases have been connected by a black line.

We see that these results are consistent with observational results. When compared
to Figure 3 in V05, for example, 
the black, red and cyan points
correspond to FIR-active (starbursting) galaxies, irrespective of the
GRASIL-3D parameter set we use, while
the green points fall in the observational range for
FIR-intermediate (mild starbursts) galaxies, and the blue points in that of either FIR-intermediate or quiescent galaxies. 
 It is worth noting that, as
expected, the most FIR-active galaxy phases among those analyzed
correspond to the black symbols, that is, just at the time when
the starburst in Figure~\ref{Age_d_z0} is at the top of its star
formation activity.  
 Moreover, as the black line
highlights, the time evolution is also as expected. It
causes    a  correlation  between the $L_{\rm FIR}/L_{\rm B}$ and the
IRAS 60 $\mu$m/ 100 $\mu$m flux density ratios as
Figure~\ref{FIRoB}   clearly shows.


\begin{figure*}
\includegraphics[width=1.\textwidth, angle=0]{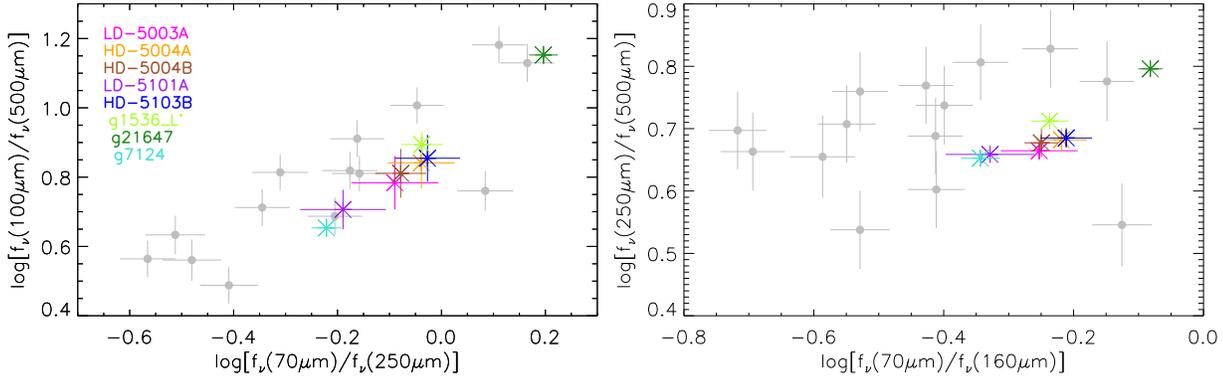}  
\caption{ Comparison of some flux density ratios in Herschel bands for the $z=0$ sample of simulated disk-like galaxies (colors)
to the corresponding ratios for the non-interacting local spiral galaxies selected by Lanz et al. 2013
 (gray symbols).  Simulated points are averages over results as model parameters take their allowed values (Table 2 for N phase), and 
error bars stand for their dispersions.
}
\label{NonI_Herschel}
\end{figure*}


\begin{figure*}
\includegraphics[width=1.0\textwidth, angle=0]{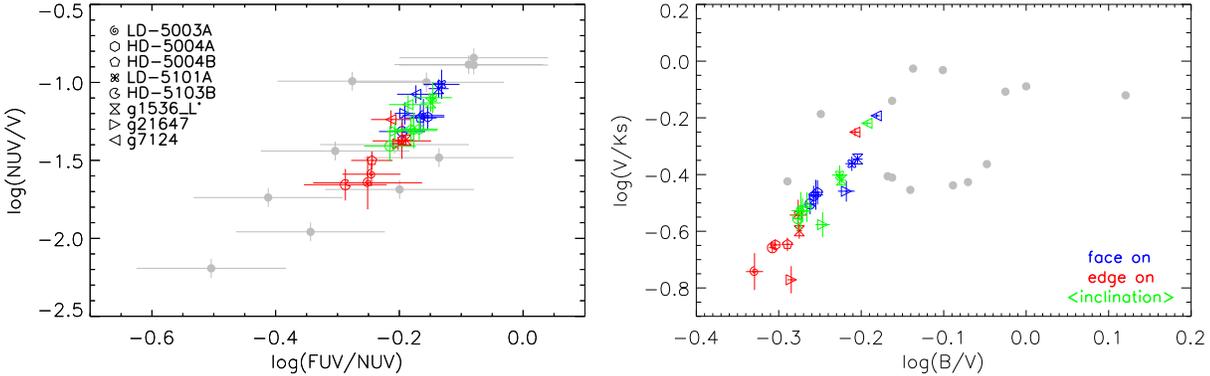}  
\caption{UV and optical flux density ratios plots for the sample of $z=0$ simulated normal disk-like galaxies averaged over parameter sets, with their corresponding dispersions, in comparison to the sample of non-interacting galaxies from Lanz et al. (2013, grey points). Data are from Dale et al. (2007), where no error bars are provided for the flux density ratios shown  in the right-hand panel. 
Colors stand for different disk orientations: face-on (blue), edge-on (red) and angle-averaged (green). 
}
\label{UVop}
\end{figure*}


To emphasize the correctness of GRASIL-3D results regarding the location of normal non-starbursting disk galaxies in this plot,
results for the whole sample of simulated $z=0$ disk galaxies described
in Section~\ref{DiskGal}  are shown in  Figure~\ref{FIRoB} upper right-hand panel. In this case, colors distinguish
galaxy identities, as specified in the left of the panel,
while symbol shapes mean different parameter Sets, as above. 
We see that these non-starbursting disk-like galaxies
show flux density ratios characteristic of FIR-quiescent or intermediate galaxies, according to V05 classification. 

To further probe the capability of GRASIL-3D to describe starbursting against more quiescent galaxies,
in Figure~\ref{FIRoB}, lower panel on the left, the f$_{\nu}$(6.75$\mu$m)/f$_{\nu}$(15$\mu$m)  flux ratios are plotted against f$_{\nu}$(60$\mu$m)/f$_{\nu}$(100$\mu$m).
Symbol shapes or colors
have the same meaning as in the upper left-hand panel.
This plot can be directly compared with Figure 4 of V05. 
A good consistency of simulations with the observational data can be appreciated regarding the location of the quiescent  $z=0$ phase and the starbursting ones
relative to observational points. 
Moreover, when the different snapshots around the HD-5103B galaxy merger are
considered, the motion of the color-color representative points
in the plane  follows  trajectories (i.e., the black line) consistent with those shown
in Figure 4 of V05, once we take into account the much
higher complexity of the merger for simulated galaxies as
compared to the models plotted in V05
\footnote{For example, in the case of simulated galaxies, the merger causes different SF
mixed episodes, the gas and molecular cloud content are local
and variable along the merger, and the optical depth for MCs is
also local and variable in time.}.

Figure~\ref{FIRoB}, lower panel on the right, shows the same flux ratios as shown in the left, in this case for the entire sample of simulated $z=0$ disk-like galaxies. 
We see that the consistency is mostly good relative to non-starbursting galaxies in Dale (2000) sample.
 
Summing up, the four panels in Figures~\ref{FIRoB} demostrate that GRASIL-3D is a good tool to calculate the SED of galaxies in different phases of their SF activity, 
including  the wet merger sequence causing the starbursts. Indeed, its capability in these tasks is comparable to GRASIL.

As mentioned above ($\S$\ref{ObsData} point (ii)), \citet{Lu:2003} provide ISOPHOT spectra for a sample of 45
disk galaxies from the same
ISO Key
Project on Normal Galaxies, providing their average
stacked spectra. They analyze the AFEs  at 6.2, 7.7, 8.6, and 11.3 $\mu$m dominating
the mid-infrared (MIR). Note the remarkable
similarity between their average rest-frame spectra
\citep[][Figure 3]{Lu:2003}  and those shown in
Figure~\ref{Disk_51.50} upper panel on the left and its $F_{\nu}$ zoomed version.

Therefore, another possible interesting comparison with observational results is
provided by the relative strengths of the rest-frame AFEs for
this same sample (Lu et al. 2003, Figure 8 and Table 6). For
the five snapshots analyzed in the HD-5103B galaxy they are shown in Figure~\ref{AFEratio}. The
agreement is good and 
it supports  these authors'
main claim, namely that the dispersion in the AFE ratios is low
for their sample (in the case of simulated results, this is
particularly true when GRASIL-3D parameter sets are
individually followed), and that little correlation is seen
between their variations and either the IRAS f$_{\nu}$(60$\mu$m)/f$_{\nu}$(100$\mu$m)  flux density
ratios or the FIR/blue luminosity ratios (i.e., galaxy SF phase activity).
Not shown in the Figure are the relative strengths of the AFEs corresponding to the
entire z=0 sample of normal simulated disk-like galaxies, these relative strengths
are compatible with those corresponding to FIR-quiescent and intermediate observed galaxies.

The galaxies in the {\it Spitzer} Infrared Nearby Galaxies Survey \citep[SINGS, see][and point (iii) in $\S$\ref{ObsData}]{Dale:2005, Dale:2007} constitute an
essential reference sample to compare with the IR emission of any calculated SEDs. 
In this case, a subsample  of  non-interacting galaxies is available \citep{Smith:2007}, allowing us to make comparisons separately.

We have therefore calculated the IRAC and
MIPS flux ratios at 3.6, 4.5, 5.8, 8.0, 24, 70 and 160 $\mu$m
for our sample of $z=0$ disk-like galaxies.
In calculating  magnitudes at 3.6, 4.5, 5.8, 8.0 and  24
$\mu$m, we used zero points of 280.9, 179.7, 115.0, 64.1 and
7.14 Jy, respectively (IRAC Data Manual; MIPS Data Manual).
In Figure~\ref{IRAC_fluxratio} we show the fluxes in these 5 bands. 
Points are averages over results with different  parameter Sets in 
Table~\ref{DiskParTable} marked (N), and the  flux histograms correspond to the sample
of 26 non-interacting local galaxies analyzed in \citet{Smith:2007}. 
No color corrections were applied in calculating these fluxes. 
The agreement is good, proving that GRASIL-3D  correctly returns fluxes in Spitzer filters for spiral galaxies.


A more restrictive selection of non-interacting galaxies from the normal galaxy sample of
\citet{Smith:2007} is provided by \citet{Lanz:2013}, see (iii) and (iv) in $\S$~\ref{ObsData}. 
 Figure~\ref{NonI_Herschel} shows some FIR flux density  ratio diagrams
for our sample of 8 disk-like galaxies at $z=0$ and the corresponding flux density ratios for the  
observed ones. Again we see that the consistency is satisfactory, showing that GRASIL-3D 
gives sound results in Herschel bands.

Apart from the IR analysis, colors in the UV and optical
wavelengths have also been analyzed, see Figure~\ref{UVop} where face-on, edge-on and angle-averaged
disk orientations are shown separately for simulated galaxies.
As shown by these plots, the comparison to non-interacting local spiral galaxies \citep{Lanz:2013}
is satisfactory for the sample of $z=0$ disk-like galaxies.

As already mentioned, parameter changes within their allowed ranges have effects on the galaxy SEDs as well as
on IRAS-ISO colors and AFE relative strengths. This will discussed in $\S$\ref{ParDiscu}.


\subsection{A Starburst Galaxy at High z}
\label{Merger7629}

A second very interesting potentiality  of GRASIL-3D is its
application to  high-$z$ massive (multiple)-merging  systems,
picked in the fast phase of their assembly process
\citep{DT:2006,Oser:2010,Cook:2009,DT:2011}. 
The timeliness of this application is reinforced
by the fact that new observational facilities, such as Herschel and ALMA, 
observe at wavelengths where the IR emission maxima of these objects are shifted to, 
providing us with observational data to compare with,
see for example \citet{Dowell:2013}, \citet{Hodge:2013} and \citet{Decar:2013}, among others.
 
One example of
such systems is provided by the \#7629 P-DEVA  simulation,
specifically designed to check GRASIL-3D when applied to such
systems. In this simulation a box of 10 Mpc side has been
sampled with $N_{DM} = 256^3$ DM and $N_{bar} = 256^3$  gas
particles, and evolved using a  spatial resolution of $\epsilon
= 400/h$ pc. The cosmological model corresponds to a flat
$\Lambda$CDM with $\Omega_{\Lambda}$ = 0.72,  $\Omega_{M} =
0.28$, $\Omega_{bar} = 0.046$ and $h=0.7$ (very similar to the
cosmological parameters used in the disk runs). To have
massive enough systems at high $z$s, a normalization of the
initial perturbation field higher than usual has been used
($\sigma_{pert} = 1.2$), therefore representing a dense
subvolume of the universe, and the  SF  parameters  have been
taken as $\rho_{*} = 3 \times 10^{-25}$ g cm$^{-3}$ and $c_{*}
= 0.1$.

Different massive objects at high $z$s have been identified in
the \#7629 simulation. In this paper we focus on one of them labeled as  D-6254.
Its  age distribution is given in Figure~\ref{sfr7629}, 
showing that the object  is a strong starburst. In Table~\ref{M7629ParTable}
we give the GRASIL-3D parameter sets used in the SED analysis.
Note that, again, those governing the molecular gas mass
fraction  $f_{mc}$ explore the entire range of their possible
values, while $t_0$ and parameters setting the properties of
individual MCs take values consistent with those
S98 found for local starbursting galaxies (ARP220).

 The corresponding gas mass and baryon fractions, $f_{mc}$ and
$f_{mc, star}$ respectively, can be found in
Table~\ref{MCresults} where we see that, again, the mass in
molecular clouds increases from parameter Sets~SB5, SB8 and SB18
 to Sets~SB4, SB7 and SB17, as $\sigma$ increases from 2 to 3,  and increases further
 for Sets~SB6, SB9 and SB19, where  $\rho_{mc, thres}$ decreases from  
3.3$\times 10^9$  M$_{\odot}$kp$^{-3}$ to  3.3$\times 10^8$   M$_{\odot}$kp$^{-3}$.
 Moreover, if the conversion factor from the observed CO
line flux to molecular gas mass is taken $\alpha \sim 1$
instead of $4.36$, as likely needed for starburst galaxies \citep{Tacconi:2013},
 then the values we found for the parameter
Sets  SB4 (SB7 and SB17) and SB6 (SB9 and SB19) are consistent with those found by these authors
for massive galaxies between redshifts $z \sim 2.0 - 2.5$,
the most distant where such analysis has been made  so far.  The
molecular gas content corresponding to the parameter Set~SB5
(as well as Sets SB8 and SB18) are too low.

The rest-frame SED of this object calculated with parameter Set SB7, 
as well as its  zoom   in the AFE  region, are shown in
 Figures~\ref{Disk_51.50}, bottom left- and right-hand panels, respectively.
 These two Figures illustrate that, as expected in a strong starburst object, the dust emission is dominated by
MCs at any IR  wavelength, and that, moreover, dust  emission clearly dominates over the extinguished stellar emission at shorter wavelengths.

\section{The effects of model parameters on the SEDs of simulated objects}
\label{ParDiscu}


\begin{table}
\centering \caption{Different parameter sets used to analyze the effects of parameter variation on the D-6254 galaxy SEDs}
\begin{tabular}{c ccc}
 \hline
Set  &   $t_0$ &  $\rho_{mc, thres}$   & $\sigma$     \\
     &   (Myrs)&  (M$_{\odot}$kp$^{-3}$) &           \\
\hline
\hline
& & $r_{mc}=$10.6 pc &   \\
\hline
\hline
  SB4       & 20      &      3.3$\times 10^9$          &  3          \\
 SB5        & 20      &      3.3$\times 10^9$          &  2      \\ 
 SB6         & 20      &      3.3$\times 10^8$          &  3       \\ 
\hline
  SB7       & 40      &      3.3$\times 10^9$          &  3    \\ 
 SB8        & 40      &      3.3$\times 10^9$          &  2    \\ 
 SB9         & 40      &      3.3$\times 10^8$          &  3    \\ 
\hline
\hline
& & $r_{mc}=$17 pc  &  \\
\hline
\hline
SB17        & 40      &      3.3$\times 10^9$          &  3     \\ 
SB18        & 40      &      3.3$\times 10^9$          &  2     \\ 
SB19        & 40      &      3.3$\times 10^8$          &  3     \\ 
\hline
\hline
\end{tabular}
\label{M7629ParTable}
\end{table}


An important question is how the change of parameters in the
model affects the SEDs of the different simulated objects we
consider. To analyze this point, for each object different SEDs
have been calculated by changing these parameters within their
allowed ranges, see $\S$\ref{Parameters} and 
Tables~\ref{DiskParTable}  and \ref{M7629ParTable}.


\begin{table*}
\centering \caption {Stellar masses, luminosities and energy balances (in parenthesis) of the HD-5103B galaxy at
$z=0$  according to the different parameter sets in Table 2   }
\begin{tabular}{c ccccc}
 \hline
Set code & Young Stellar & Free Stellar  & $L_{bol}$ &  $L_{c}$ & $L_{mc}$ \\
& ($10^{6} M_{\odot}$) & ($10^{10} M_{\odot}$)   & ($10^{44}$ erg sec$^{-1}$) & ($10^{44}$ erg sec$^{-1}$) & ($10^{44}$ erg sec$^{-1}$)  \\
\hline
\hline
 1  &  4.9136 & 2.5363  & 0.7871 (98.9) & 0.2524 (99.56) & 0.1164 (96.7) \\
 2  &  4.9136 & 2.5363  & 0.7783 (97.8) & 0.3172 (99.8)  & 0.1140 (98.8) \\
 3  &  4.9136 & 2.5363  & 0.7934 (99.7) & 0.1629 (99.0)  & 0.1165 (96.6) \\
\hline
  4  &  6.0475 & 2.5362  & 0.7729 (99.5) & 0.2334 (99.3) & 0.1276 (96.9)    \\
  5  &  6.0475 & 2.5362  & 0.7665 (98.7) & 0.3003 (98.4) & 0.1252 (98.8)   \\
  6  &  6.0475 & 2.5362  & 0.7777 (99.9) & 0.1462 (99.6) & 0.1285 (96.3)  \\
\hline
  7  & 14.741 & 2.5353  & 0.7723 (99.5) & 0.2094 (100.0) & 0.1649 (97.9)   \\
  8  & 14.741 & 2.5353  & 0.7644 (98.4) & 0.2740 (98.7) & 0.1602 (99.2)    \\
  9  & 14.741 & 2.5353  & 0.7767 (99.9) & 0.1270 (98.6) & 0.1655 (97.6)    \\
\hline
\hline
  13  & 14.741 & 2.5353 &  0.7780 (99.8) & 0.2103 (99.9) & 0.1703 (94.8)  \\
  14  & 14.741 & 2.5353 &  0.7730 (99.6) & 0.2759 (98.6) & 0.1686 (95.8) \\
  15  & 14.741 & 2.5353 &  0.7826 (99.2) & 0.1274 (98.7) & 0.1713 (94.2) \\
\hline
\hline
\end{tabular}
\label{PropTab5103}
\end{table*}


To properly interpret the results shown in
Table~\ref{PropTab5103}, we recall that the total gas or
stellar mass of a simulated galaxy, as well as its total
bolometric luminosity in the absence of dust, are fixed by the
simulation itself. Therefore, the sum of MC and cirrus total
masses is fixed to the total gas mass, and the sum of the
young and free stars is also fixed to the galaxy total stellar
mass.


\begin{figure*}
\centering
\includegraphics[width=.35\textwidth, angle=270]{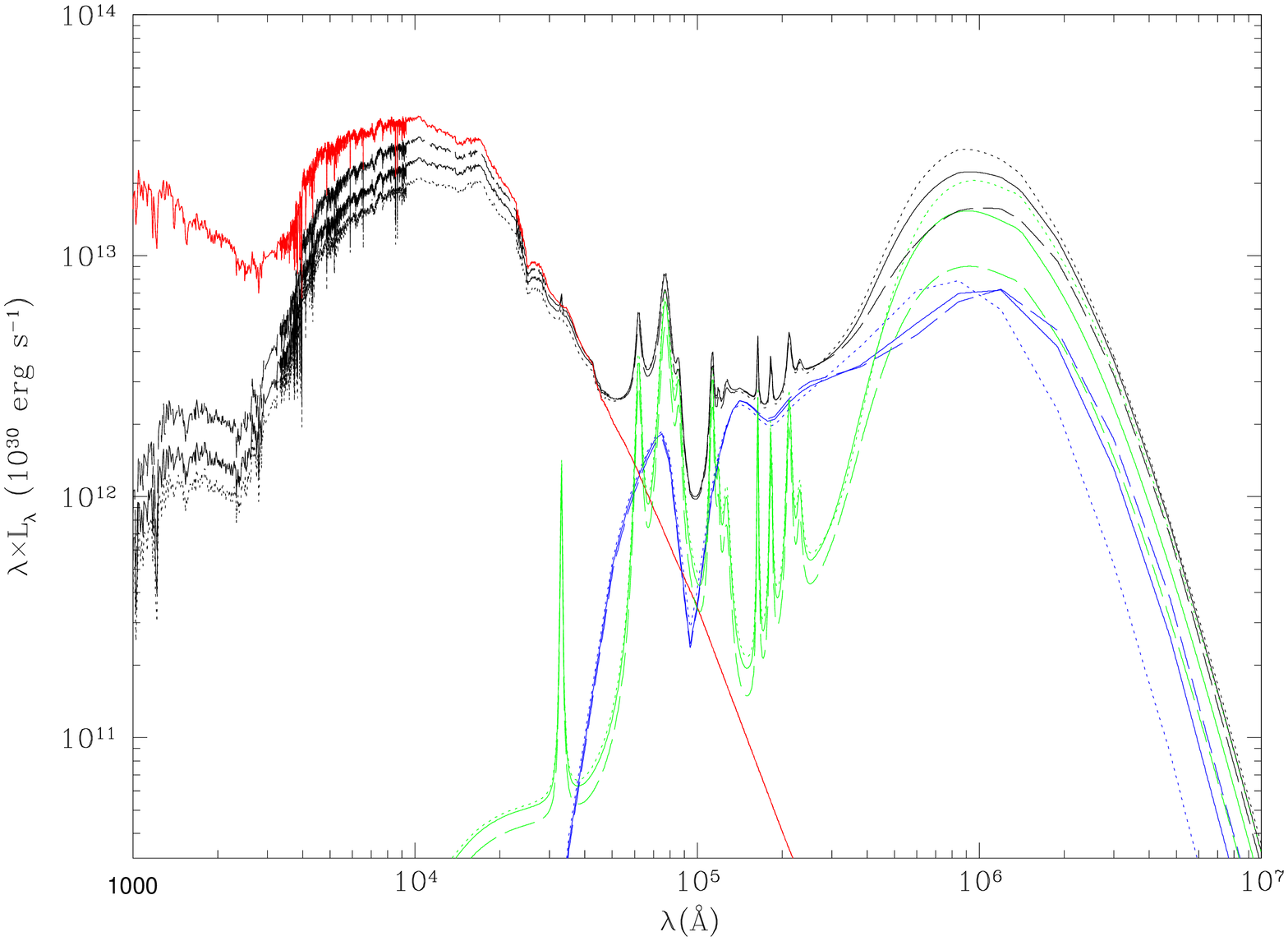}   
\includegraphics[width=.35\textwidth, angle=270]{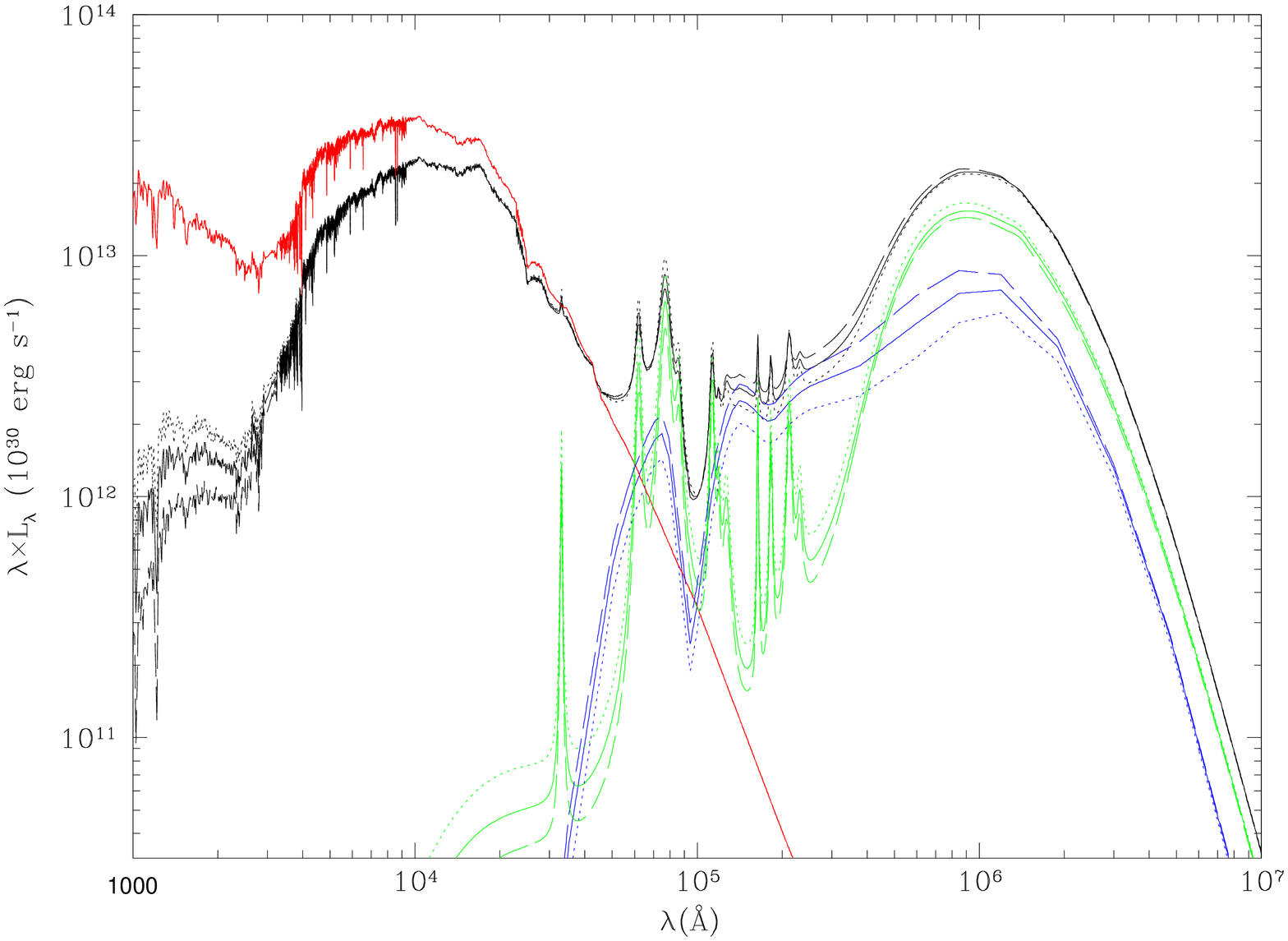}     
\caption{Comparison of the SEDs of the simulated galaxy HD-5103B
at $z=0$ obtained under different parameter sets. Left-hand panel
compares the effects of changing $\rho_{mc, thres}$ and
$\sigma$, with full, point and dashed lines corresponding to
code 4, 5 and 6 parameter Sets. Right-hand panel compares the
effects of changing the $t_0$ parameter, with full, point and
dashed lines corresponding to code 4, 1 and 7 parameter Sets.
See Table 2 for parameter set values. Color line codes are as
usual and only angle-averaged emissions are plotted. } \label{ParComp}
\end{figure*}


\subsection{The SEDs}

As an illustration of the effects of parameter variations on
the SED, in Figures~\ref{ParComp} we show the SED of HD-5103B
galaxy at $z=0$,  calculated using parameter Sets 4, 5 and 6
(i.e., different $\rho_{mc, thres}$ and $\sigma$
values, left-hand panel) and parameter Sets 4, 1 and 7
 (i.e., different $t_0$ values, right-hand panel). On the left-hand panel we see
that increasing $f_{mc}$  (i.e., decreasing the cirrus fraction
5-4-6 parameter Sets), the dust absorption by cirrus in the UV and
optical decreases, thereby decreasing the cirrus emission at longer
wavelengths. The right-hand panel illustrates the consequences of
increasing the time young stars remain within MCs: as $t_0$
increases from parameter Sets  
1, 4, 7, the UV emission
decreases, but the MC cloud emission increases, at the same
time that the cirrus emission in the PAH region decreases.

\subsection{The global masses and luminosities of the different galaxy components}

As expected, Tables~\ref{MCresults} and \ref{PropTab5103} show
that the global masses of the different galaxy  components are
insensitive to $m_{mc}$ and $r_{mc}$ variations, and the global
luminosities $L_{bol}$\footnote{Note that the values of the
bolometric luminosity, $L_{bol}$, given in the the fourth
column of Table~\ref{PropTab5103} are constant and equal to the
intrinsic stellar bolometric luminosity of HD-5103B, within the
accuracy of the calculation (shown by the number in parenthesis
at their right).} and  $L_{c}$ do not change.
There is a tendency  towards slightly higher
values of  $L_{mc}$  as $\tau_{mc}$ decreases.


\begin{table*}
\begin{minipage}{6.5in}
\renewcommand{\thefootnote}{\thempfootnote}
\centering \caption{Galaxy HD-5103B at different $z$s,
g1536\_L$^{*}$ at $z=0$ and D-6254 at $z=4$: Molecular Cloud and Cirrus Dust Content
according to $\rho_{mc, thres}$ and  $\sigma$ }
\begin{tabular}{c ccccc}
 \hline
Group & Set  &  Molecular Clouds &  $f_{mc}$\footnote{$f_{mc} = M_{mc}/M_{\rm gas}$} &    $f_{mc, star}$\footnote{$f_{mc, star} = M_{mc}/(M_{mc} + M_{star})$ as in Tacconi et al. 2013}    &  Dust in Cirrus    \\
    & &   (10$^{10}$ M$_{\odot}$) &      &               & (10$^{7}$ M$_{\odot}$)         \\
\hline
\hline
     &    &            &    HD-5103B                                 &                          & \\
\hline
\hline
      &   & $t/t_U$ = 1 & $M_{\rm tot}$\footnote{$M_{\rm tot} = M_{\rm star} + M_{\rm gas}$} =2.973 $\times 10^{10} $ M$_{\odot}$ &  $M_{\rm gas} = 4.362 \times 10^{9}  $ M$_{\odot}$ \\
\hline
1 & 1, 4, 7, 13  &  0.1433 &  0.3286  & 0.0460  &  4.0385   \\
2 & 2, 5, 8, 14  &  0.0257  &  0.0590  & 0.0100  &  5.8017    \\
3 & 3, 6, 9, 15  &  0.2598   &  0.5956  & 0.0929  &  2.3493  \\
\hline
     &   & $t/t_U$ = 0.394 & $M_{\rm tot} =1.397 \times 10^{10} $ M$_{\odot}$ & $M_{\rm gas} = 3.762 \times 10^{9}  $ M$_{\odot}$ \\
\hline
1 & 7, 10, 13  & 0.1674  & 0.4449 & 0.1409  & 1.1386 \\
2 & 8, 11, 14  & 0.0855  & 0.2273 & 0.0773  & 1.9336 \\
3 & 9, 12, 15  & 0.2423  & 0.6440 & 0.1918  & 0.5516  \\
\hline
      &  & $t/t_U$ = 0.377 & $M_{\rm tot} =1.357 \times 10^{10}  $ M$_{\odot}$ & $M_{\rm gas} =  3.771 \times 10^{9} $ M$_{\odot}$ \\
\hline
1 & 7, 10, 13  & 0.1981 & 0.5253 & 0.1682 & 1.1156 \\
2 & 8, 11, 14  & 0.1297 & 0.3439 & 0.1169 & 1.8727 \\
3 & 9, 12, 15  & 0.2596 & 0.6886 & 0.2095 & 0.5573 \\
\hline
      &  & $t/t_U$ = 0.361& $M_{\rm tot} =1.295 \times 10^{10}  $ M$_{\odot}$  & $M_{\rm gas} = 4.369 \times 10^9  $ M$_{\odot}$  \\
\hline
1 & 7, 10, 13  & 0.2773 &  0.6346  & 0.2441 & 0.9128 \\
2 & 8, 11, 14  & 0.2127 &  0.4868  & 0.1986 & 1.5082 \\
3 & 9, 12, 15  & 0.3328 &  0.7617  & 0.2794 & 0.4631    \\
\hline
      &  & $t/t_U$ = 0.344& $M_{\rm tot} =1.260 \times 10^{10}   $ M$_{\odot}$  &  $M_{\rm gas} = 5.586 \times 10^9 $ M$_{\odot}$ \\
\hline
1 & 7, 10, 13  & 0.3351  &  0.5998 & 0.3252  & 1.3797 \\
2 & 8, 11, 14  & 0.1761  &  0.3153 & 0.2009  & 2.6681 \\
3 & 9, 12, 15  & 0.4363  &  0.7811 & 0.3834  & 0.5958    \\
\hline
\hline
       &  &            &    g1536\_L$^{*}$                                &                          & \\
\hline
\hline
      &   & $t/t_U$ = 1 & $M_{\rm tot} = 3.317 \times 10^{10} $ M$_{\odot}$ & $M_{\rm gas} = 1.082 \times 10^{10}  $ M$_{\odot}$ \\
\hline
1 & 1, 4, 7,  13  &   0.1750  &  0.1617 & 0.0726 & 6.1449 \\
2 & 2, 5, 8,  14  &   0.0109  &  0.0101 & 0.0049 & 7.3175 \\
3 & 3, 6, 9,  15  &   0.4039  &  0.3731 & 0.1530 & 4.5585 \\

\hline
\hline
      &   &            &    D-6254                               &                          & \\
\hline
\hline
     &   & $t/t_U$ = 0.110 & $M_{\rm tot} = 3.298 \times 10^{10}  $ M$_{\odot}$ & $M_{\rm gas} = 7.293 \times 10^9  $ M$_{\odot}$ \\
\hline
1 & SB4, SB7, SB17         & 0.3286 & 0.4506 &  0.1134              &   1.2033     \\
2 & SB5, SB8, SB18         & 0.1228 & 0.1684 &  0.0456              &   2.2805     \\
3 & SB6, SB9, SB19        & 0.4856 & 0.6656 &  0.1590              &   0.5395        \\
\hline
\hline

\end{tabular}
\label{MCresults}
\end{minipage}
\end{table*}


The Tables  also show that when $t_0$ increases, the global
mass of young stars suffers an important increment in relative
terms. The global mass of free stars decreases, but the effect
is relatively less important. As a consequence, the total
stellar energy to be processed by MCs increases, and, at the
same time, the amount of stellar energy that directly heats the
cirrus component becomes relatively less important. The
resulting effect is that $L_{mc}$ increases, while $L_{c}$
decreases.

Regarding changes in $\rho_{mc, thres}$ and $\sigma$, we treat
them in turn. When $\rho_{mc, thres}$ decreases, the global
mass in MCs  (and $f_{mc}$) increases, causing a decrement to
the global gaseous mass in cirrus and to their dust
content. A direct effect is that $L_{c}$ decreases. $L_{mc}$
keeps roughly constant, which can be understood as a
consequence of the constant  amount of energy from young
stars heating the MC component.

The effects of $\sigma$ variations on the global masses and
$L_{c}$ and $L_{mc}$ luminosities are in the opposite direction
of those caused by $\rho_{mc, thres}$ changes, as can be seen
in Tables~\ref{MCresults} and \ref{PropTab5103}. When both
parameters vary, the effects can be somewhat compensated.

\subsection{Flux density ratios}

 In this case, the major effects come from $f_{mc}$
value changes due to $\rho_{mc, thres}$ and $\sigma$
modifications, as illustrated by Figure~\ref{ParComp}. These
effects are more apparent in the FIR/blue luminosity
ratios shown in Figure~\ref{FIRoB}, upper panels. 
Indeed, results corresponding to
parameter sets bringing about degenerated   $f_{mc}$  values,
namely \# (2, 5, 8, 11, 14; open symbols), \# (1, 4, 7, 10, 13; filled symbols), and 
\# (3, 6, 9, 12, 15; composed symbols), show up grouped together in FIR/blue luminosity ratios.
According to  Table~\ref{MCresults},
these groups of parameter sets correspond to increasing molecular gas content,
and hence, to increasing L$_B$ and decreasing cirrus emission in
the FIR region. Indeed, in Figure~\ref{FIRoB} (top panels) we see that parameter Sets 2, 5, 8,
11 and 14 (not considered for {\tt GASOLINE} galaxies)
 have the highest FIR/blue luminosity
ratios. 
The effects are more
important for FIR-active   galaxies than for FIR-quiet
phases, and they are within the data dispersion at given
f$_{\nu}$(60$\mu$m)/f$_{\nu}$(100$\mu$m) ratios, that are much less affected by $f_{mc}$
variations.

The effects of changing the timescale for MC destruction,
$t_0$, are not important on the FIR/blue luminosity ratios,
$L_{\rm FIR}/L_{\rm B}$, and only modest on the IRAS 60 $\mu$m/
100 $\mu$m flux density ratios, see Figure~\ref{FIRoB}. 
Finally, changing the size of individual MCs
corresponding to a 50\% variation in their optical depth
(Sets 13, 14 and 15 versus 7, 8 and 9), has no appreciable
effects in Figure~\ref{FIRoB}.

Other flux density ratios affected by changes in $f_{mc}$ 
are the IRAS-ISO  6.75 $\mu$m/ 14 $\mu$m 
 flux ratios (Figure~\ref{FIRoB} lower panel) as well as the
FUV/NUV and  V/Ks  flux
density ratios, effects shown as error bars in Figure~\ref{UVop}. Note however that
the effects are consistent with the corresponding
data dispersion. In this
respect, we remind that $f_{mc}$ changes due to parameter
variations are generally consistent with observational data on $f_{mc}$.
In the case of parameter Set 5 (and 2, 8, and 14) the
corresponding points in Figures~\ref{HI_H2} are at the limits
of the data cloud of points or outside it.

Finally, it is remarkable to note that, as shown by
Figures~\ref{FIRoB} and  \ref{AFEratio}, the effects of
modifying the $t_0$ or $r_{mc}$ parameters within their allowed
ranges has insignificant effects in the disk flux density ratios
analyzed in this paper.

  These detailed analyses have been repeated for
 the   D-6254 high-$z$ merger galaxy by comparing results obtained
under parameter sets in  Table~\ref{M7629ParTable}, and the conclusions
are similar to those reached for disk galaxies.

\section{Summary, Discussion and Conclusions }
\label{SummConclu}

In this  paper we have introduced a new photometric SED code,
GRASIL-3D,  which includes a careful modelling of the radiative
transfer through the dust component of the ISM. GRASIL-3D is an
entirely new model based on the formulism of an existing and widely applied model,
GRASIL, but specifically designed to be applied to systems with
any arbitrarily given geometry, where radiative transfer through dust plays an
important role, such as galaxies calculated by hydrodynamical
galaxy formation codes.

A few codes exist that, by interfacing the outputs of
hydrodynamic simulations, can Monte Carlo solve the radiative transfer through
dust and therefore predict a multi-wavelength SED for simulated
galaxies (SUNRISE, RADISHE, ART$^{2}$, see $\S$\ref{intro}).
Following GRASIL, some GRASIL-3D particular strengths relative
to these codes can be summarized as: i) the radiative transfer is not solved
through Monte Carlo methods, but in a grid; ii) it is designed
to separately treat radiative transfer in molecular clouds and
in diffuse cirrus component, whose dust composition are
different (for example, lower PAH fraction in MC dust); iii) correspondingly, it takes
into account the age-dependent dust reprocessing of stellar
populations (note that GRASIL has been the first model to do
so), arising from the fact that younger stars are associated
with denser ISM environments, mimicking through the $t_0$
parameter the time young stars are enshrouded within molecular
clouds before their destruction; iv) it includes a detailed
non-equilibrium calculation  for dust grains with diameter
smaller than $a_{flu} \sim 250$ \AA, as required. This allows
also a proper treatment of PAH features, dominating the MIR in
some cases.

Current cosmological hydrodynamical codes that follow galaxy
formation cannot resolve molecular clouds, therefore some further
modelling is required. A sub-resolution model to calculate the
local mass in the form of molecular clouds has been implemented
in GRASIL-3D, based on a theoretical log-normal PDF for the gas
densities, as suggested by small scale ($\sim$ 1 kpc)
simulations, and on the assumption that MCs are defined by a
threshold, $\rho_{mc, thres}$. In this model, the mass of MC
relative to the total mass gas, $f_{mc} = M_{mc}/M_{gas}$,
 as well as the cirrus density field, is
set by two parameters, $\rho_{mc, thres}$ and $\sigma$, the PDF
dispersion. The entire range of values for these parameters
given by the literature has been explored in GRASIL-3D, with the
result that $f_{mc}$ is consistent with observations, with the  exception of 
three galaxy models and a set of extreme parameter values
producing a value of $f_{mc}$ which is too low. This is a very important
consistency check for GRASIL-3D.

When solving the radiative transfer for cirrus, a concern is in order when
implementing the calculation of the radiation field in a grid
cell due to the emissions of this same cell (where an apparent
singularity appears), or the absorption of radiation emitted at
the $k$-th  cell along a given direction within this same cell. In
these cases, at high optical thicknesses, representing the cell
by its central point gives a poor representation of the
physical processes resulting in inaccurate energy balances. An
elegant improvement has been implemented in GRASIL-3D, where
the cells are Monte Carlo split into $N_p$ points representing
sub-volumes, and the calculations made based on these
decompositions. As a result, energy balances improve
considerably, being $\sim 98\% $ or higher in most cases.

The code has a general applicability to the outputs of
simulated galaxies, either using Lagrangian or Eulerian
hydrodynamic codes. As an application, the  new model has been
interfaced  with the   P-DEVA    and   {\tt
GASOLINE} SPH codes. As first applications, and to show the code
potentialities,  GRASIL-3D has been used to calculate  the SEDs
for a variety of simulated  galaxies: a sample of 8 normal 
non-interacting disk-like galaxies at $z=0$, a
merger event between disks, and a high-$z$ massive galaxy in the
phase of its fast  mass assembly.

A detailed analysis of the calculated SEDs has been performed,
putting  particular emphasis  in the rest-frame mid to far-IR
region, where the effects of cirrus and MC emission dominate.
Comparisons with data coming from different projects have been
discussed for the disks, finding very encouraging results.
Particularly remarkable are the agreements related to the PAH
features, a very important SF discriminator, thereby opening
interesting possibilities for applications of the code. In particular,
GRASIL-3D allows the creation of  2D images of such galaxies, at
several angles and at different bands from UV to sub-mm.

The consequences of GRASIL-3D parameter variations on the SEDs
of galaxies has been analyzed in detail, and the results point
to no remarkable effects when parameter variations are kept
within their allowed ranges. The main effects come from
$f_{mc}$ variations due to $\rho_{mc, thres}$ and $\sigma$
modifications. 
In this case, the effects are not particularly  relevant when
compared to data dispersion, as long as $f_{mc}$ values are
compatible with recent data on molecular gas content of
galaxies \citep{Saintonge:2011, Saintonge:2012, Tacconi:2013}.

In forthcoming papers these applications will be analyzed in
more detail and in a wider perspective.

\section*{Acknowledgments}

 We thank Mariola Dom\'enech, Fran Mart\'{\i}nez-Serrano and
Greg Stinson for allowing us to use results of simulations. 
We also thank the anonymous referee whose constructive criticisms and recommendations
have helped to improve this paper.
This work was partially supported by the MICINN and MINECO (Spain) through the grants
AYA2009-12792-C03-02 and AYA2009-12792-C03-03 from the PNAyA, as well as by
the regional Madrid V PRICIT program through the ASTROMADRID network
(CAM S2009/ESP-1496) and the ''Supercomputaci\'on y e-Ciencia''
Consolider-Ingenio CSD2007-0050 project.
We also thank the computer resources provided by BSC/RES (Spain)
and the Centro de Computaci\'on Cientif\'ica (UAM, Spain).
 P. Alpresa and  A. Obreja thank the MICINN and MINECO (Spain) for financial
support through  FPI fellowships.
C. Brook thanks MINECO for financial support through contract associated 
to AYA2009-12792-C03-03 grant.


\bibliographystyle{mn2e}
\bibliography{rdt_V4}


\bsp

\label{lastpage}

\end{document}